\documentclass[useAMS,usenatbib]{mn2e} 
\usepackage{epsfig}

\def\spose#1{\hbox to 0pt{#1\hss}} 
\def\lta{\mathrel{\spose{\lower 3pt\hbox{$\mathchar"218$}}      
     \raise 2.0pt\hbox{$\mathchar"13C$}}}      
\def\gta{\mathrel{\spose{\lower 3pt\hbox{$\mathchar"218$}}      
     \raise 2.0pt\hbox{$\mathchar"13E$}}}      
   
\title[The Chemical Evolution of Globular Clusters]{The Chemical Evolution 
of Globular Clusters - I. Reactive Elements and Non-Metals}

\author[A. Marcolini et~al.] {
       A. Marcolini$^{1}$,  B.~K. Gibson$^1$, A.~I. Karakas$^2$, P. 
	S\'anchez-Bl\'azquez$^{1,3}$ \\
       $^1$Jeremiah Horrocks Institute for Astrophysics \& Supercomputing, 
       University of Central Lancashire,
       Preston, PR1~2HE, UK \\
       $^2$Research School of Astronomy \& Astrophysics, Mt Stromlo Observatory,
       Weston Creek ACT 2611, Australia\\
       $^2$Instituto de Astrof\'{\i}sica de Canaria; c/V\'{\i}a Lactea s/n, E38205, La Laguna (Tenerife), Spain. \\                                           } 
\date{Accepted ..., Received ...; in original ...}     
   
\pagerange{\pageref{firstpage}--\pageref{lastpage}}
\pubyear{2009}
 
\begin{document}

\maketitle

\label{firstpage}

\begin{abstract}
We propose a new chemical evolution model aimed at explaining the
chemical properties of globular clusters (GC) stars.
Our model depends upon the existence of (i) a peculiar
pre-enrichment phase in the GC's parent galaxy associated with very
low-metallicity Type~II supernovae (SNe~II), and (ii) localized
inhomogeneous enrichment from a single Type~Ia supernova (SNe~Ia) and
intermediate-mass (4-7~M$_{\odot}$) asymptotic giant branch (AGB)
field stars.  GC formation is then assumed to take place within
this chemically-peculiar region.  Thus, in our model the first
low-mass GC stars to form are those with peculiar abundances (i.e.,
O-depleted and Na-enhanced) while ``normal'' stars (i.e., O-rich and
Na-depleted) are formed in a second stage when self-pollution from
SNe~II occurs and the peculiar pollution from the previous phase is
dispersed.  In this study, we focus on three different GCs: NGCs~6752,
NGC 6205 (M~13) and NGC 2808. We demonstrate that, within this framework, a
model can be constructed which is consistent with (i) the
elemental abundance anti-correlations, (ii) isotopic abundance
patterns, and (iii) the extreme [O/Fe] values observed in NGC~2808 and
M~13, without violating the global constraints of approximately
unimodal [Fe/H] and C+N+O.
\end{abstract}


\section{Introduction} 
\label{sec:introduction} 

As laboratories for both stellar and galactic physics, globular
clusters (GCs) rank among the valuable tools available to
astronomers. Ancient, co-eval, equidistant, and apparently
mono-metallic, GCs give the appearance of being elegantly simple
aggregates of stars. Behind this ``mask of simplicity'' though, lurks
perplexing observations which have defied explanation for several
decades.

First, each well-studied cluster to date shows star-to-star abundance
variations of the light elements C, N, O, Na, Mg, and Al \citep[][and
references therein]{kraft1994,gratton2004}, not shared by field stars
with similar metallicities \citep{gratton2000}. These variations
follow a common pattern with C-N, O-Na, and Mg-Al all anti-correlated
\citep[e.g][]{kraft1993,grundahl2002, cohen2002, yong2003, sneden2004,
yong2005, cohen2005, carretta2006, gratton2007, marino2008}.
Importantly, the sum of C$+$N$+$O is observed to be approximately
constant \citep[e.g.][]{ivans1999,carretta2005}. Moreover,
\citet{smith2005} noted that stars in one of the Galactic globular
clusters, NGC~6121, showed evidence for a Na-F anti-correlation,
whereas \citet{pasquini2005} and \citet{bonifacio2007} found a Na-Li
anti-correlation in NGC~6752 and 47~Tucanae, respectively.  The
abundances of Si, Ca, the iron-peak (e.g., Fe, Ni, and Cu) and
neutron-capture elements (e.g., Ba, La, Eu) do not show the same
star-to-star scatter, nor do they vary with the light elements
\citep{gratton2004,james2004,yong2006,yong2008}. The most important
exception to this is the most massive cluster $\omega$ Centauri, that
shows both, a spread in age and [Fe/H], and  a rise in s-process
element abundances with increasing [Fe/H]. Evidence for $\omega$ Cen
suggests it may have an extragalactic origin
\citep[e.g.][]{majewski2000, smith2000, gnedin2002, bekki2003,
romano2007, marcolini2007}.

The above abundance trends have been found for stars in all
evolutionary phases, from the main-sequence turn-off through to
the tip of the first giant branch, which has lent support to 
the ``self-pollution'' hypothesis \citep{cottrell1981, dantona1983}.
This is in contrast to deep mixing,  where the abundance anomalies 
are produced by internal mixing during the ascent of the 
giant branch, after the first dredge-up
\citep{sweigart79,pin97,charbonnel1994}.
Further evidence for self-pollution comes from a lack of
variation in the O-Na and Mg-Al anti-correlations with 
luminosity as the stars ascend the first giant branch. 

According to the self-pollution scenario a previous generation 
of stars $"$contaminated$"$ the  atmospheres of stars observed today in GCs,
or provided  much of the material from which those stars formed \citep[e.g.][]{jehin1998,
parmentier1999, tsujimoto2007}.  The roughly constant [Fe/H]
abundances in a given GC led to the assumption that the polluters 
were low-metallicity, intermediate-mass asymptotic giant branch 
(AGB) stars, with masses between $\sim 4$ and 7~M$_\odot$ 
\citep{dantona2004, bekki2007}. The hot bottom burning 
experienced by these stars provides the hydrogen burning 
environment (at least qualitatively) that can alter the abundances 
of the light elements. Detailed AGB models have mostly failed to explain 
the observed abundance trends 
\citep{denissenkov2003, fenner2004, cohen2005, karakas2006},
but model uncertainties, including convection and mass loss, 
render predictions uncertain \citep{ventura2005a,ventura2005b}.
Recently,  slow-winds from rotating massive stars 
\citep{prantzos2006,smith2006,decressin2007a} have also 
been proposed as another possible source of the abundance
anomalies.

The canonical AGB enrichment scenario consists of two stages in which
the first stars to form are those pre-polluted by core collapse
supernovae (SNe~II), and have [O/Fe] $\sim$ 0.5 and [Na/Fe] $\sim
-0.1$.  After the first SNe~II have driven away the gas from the
proto-cluster, gas polluted by AGB winds starts to accumulate at the
centre of the cluster and a second generation of stars with peculiar
abundances can form. These second-generation stars will have depleted
F and O, along with enhanced in N and Na with respect to the first
generation.  Variations to this scenario include (1) the AGB
ejecta-enriched gas is also accreted on to the surface of some (e.g.,
segregated) stars \citep{parmentier1999}, and (2) the AGB pollution
comes, not only from the first generation of stars, but also from all
AGB stars of the satellite galaxy where the GC is forming
\citep{bekki2007}. The second assumption was made to allow for a larger
reservoir of material for the formation of the second generation 
of stars \citep{bekki2006}, as if all AGB pollution 
comes from the first generation of stars, is very difficult to 
explain the high fraction (as high as 50\%) of peculiar-to-normal 
stars in GCs \citep[e.g.,][]{dantona2004, dantona2005,
carretta2006, piotto2007}.  The same results can be achieved by
assuming that the GCs were much more massive in the past (by a factor
of $\sim$10-100$\times$) and lost preferentially most (90-99\%) of
their first generation of stars through tidal interactions with the
Milky Way \citep[MW; e.g.][]{dercole2008}. 

Another way to solve the problems associated with the AGB-$"$pre-pollution$"$ 
scenario is to assume that the polluters were not AGB stars
but Type Ia supernovae.  \citet{marcolini2006} suggested that the
O-depleted stars observed in dwarf spheroidal galaxies (dSphs) as well
as in the peculiar globular cluster $\omega$ Cen \citep{marcolini2007}
can be explained if those stars were born in a small iron-rich
volume {\it inhomogeneously} polluted by a single SN~Ia. The ejecta of
SNe~Ia is very rich in Fe and extremely poor in O (compared to SNe~II)
thus stars forming in such confined regions would show very low values
of [O/Fe] (usually up to $\sim -0.5$). Dwarf galaxies, however, show a
large spread (1.5~dex) of [Fe/H], hence it is not clear 
if this model is applicable to the problem of the abundance anomalies in
mono-metallic GCs.

In this paper we present a GC chemical evolution model where all of
the observed chemical peculiarities arise from the inhomogeneous
pollution of a single SN~Ia and intermediate-mass AGB
stars. This pollution is mixed with gas that was pre-enriched by very
low-metallicity SNe~II. Since each of these so-called ``polluters''
precede the formation of the GC itself, we call this type of enrichment
``external pollution'' (according to Bekki et~al. 2007) from the Milky
Way (or host galaxy). When the first stars with peculiar abundances are
forming, SNe~II start to explode, self-polluting the ISM and
dispersing the inhomogeneous region. As star formation proceeds,
the abundances of the forming stars will move from ``peculiar'' to
``normal'', where normal in this context means stars with chemical
properties typical of SNe~II enrichment.  Eventually, the SNe~II
explosions will cease any further star formation (SF) and the GC will
evolve passively. In previous studies, SN~Ia had been discarded as
polluters because the [Fe/H] is essentially constant within a given GC,
and SN~Ia are efficient producers of iron. Nevertheless, we will show
that our model is able to reproduce the observed
anti-correlations, while keeping the iron content of the newly forming
stars approximately constant.

In the following we test our scenario on a GC with a typical 
[O/Fe]-depletion pattern (on the order of $\sim -0.5$), similar to
to the well-studied cluster NGC~6752 ([Fe/H]=$-1.56$). In addition
we test our scenario on a more extreme case such as that of 
NGC~2808 ([Fe/H]=$-1.15$) and M~13 ([Fe/H]=$-1.50$) \citep{harris1996}, 
with [O/Fe]-depletions as large as 1~dex.

\section{Initial Conditions}
\label{sec:initial_conditions}

It is generally assumed that GCs were assembled during the formation
of the Milky Way halo (or their parent galaxy) even though their
chemical properties differ from those halo field stars of the same
metallicity \citep[e.g.,][]{gratton2000, freeman2002}.  It is likely
 that part, or all, of this difference can be traced to the very 
different conditions experienced by the respective systems e.g., 
deeper potential wells and the associated burst of SF.

Following this framework, we assume that a typical GC forms initially
within gas pre-enriched by field polluters such as SNe~II, but also
from SNe~Ia and AGB stars associated with (parent) halo star formation.
However, due to the early stage of the formation of the halo (with
respect to the rest of the parent/Milky Way), SNe~Ia and AGB stars are
not important in shaping the $mean$ chemical properties of the gas
unless they are polluting it inhomogeneously (i.e., polluting a small
localized region). Thus, we assume that a typical GC will form
initially in a peculiar region of the halo that was pre-polluted by
SNe~II, and in which the inhomogeneous pollution of a single SN~Ia
$and$ intermediate-mass AGB stars is added.  In our model, all the
chemical peculiarities observed in GC stars (i.e., O-depleted and
Na-rich stars) which are not shared by their field counterparts arise
due to this peculiar localized inhomogeneous pollution. 

\begin{figure}    
\begin{center}    
\psfig{figure=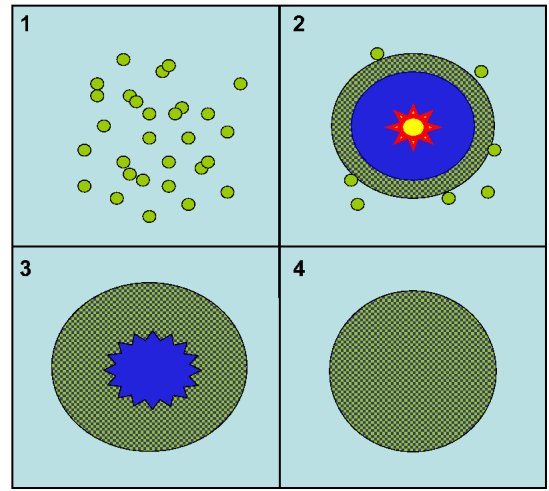,width=0.50\textwidth}
\end{center}   
\caption{Schematic framework leading to the initial conditions
required for our model: 1) low-metallicity 4--7 M$_{\odot}$ AGB stars
(green dots) start polluting a particular region of the halo which
has been pre-enriched by SNe~II; 2) a SN~Ia explodes near this region
and during its expansion collects the AGB material in its shell; 3)
the SN remnant radiates away its energy and the gas starts
re-collapsing, mixing the SN~Ia ejecta with material from the AGB
stars; 4) the final result is a peculiar region in the Galactic halo,
where on top of the mean pre-enrichment from early SNe~II, the
inhomogeneous contribution by a single SN~Ia and a number of
intermediate-mass AGB stars is added. These are our assumed initial
conditions for proto-GC formation. Note that the light-blue background
volume (mainly polluted by early-SNe~II) represents the mean chemical
properties of the gas out of which ``normal'' field halo stars are
formed.}
\label{fig:model_initial} 
\end{figure}

A schematic representation of how these initial conditions can be
achieved is shown in Fig~\ref{fig:model_initial}. According to our
model, AGB material is deposited in the vicinity of a SNe~Ia progenitor
binary system. As the secondary explodes as a SN~Ia and the supernovae
remnant (SNR) expands, it will collect this surrounding AGB material
in its shell.  The numerical evolution of a radiative SNR has been 
analyzed in detail by \citet{cioffi1988} and \citet{thornton1998}.
These studies found that for normal ambient gas density of 
$n_{0}=1-10$ cm$^{-3}$, the evolution of a SNR
lasts for a few Myrs and that it reaches radius  of
$\sim$50-90 pc (with the larger radii achieved for the smaller ambient
gas densities). After radiating away most of its energy, the remnant
will first stall, lose its identity and then collapse back
\citep[e.g.][; approximately after R$_{\rm max}/$c$_{\rm s}\sim$ 5-9
Myr, assuming an ambient sound speed c$_{\rm s}$ of 10 km
s$^{-1}$]{slavin1992}. The result of this evolution will be a localized 
region of the halo,
polluted inhomogeneously with SN~Ia and AGB ejecta.  These will be the
initial conditions of our model which we define as the ``pre-peculiar
enrichment'' scenario;  we assume that the proto-GC starts forming
stars in this peculiar region.

Regions polluted inhomogeneously by SNe Ia arise naturally in the
hydrodynamical simulations of dSphs chemical enrichment performed by
\citet{marcolini2006, marcolini2008}. \citet{cescutti2008} also showed
that a model which takes into account inhomogeneous pollution is able
to explain the observed spread in $s$- and $r$-process elements at low
metallicities in the Galactic halo \citep[see also][]{ishimaru1999,
argast2000}. Since both the peak SNe~Ia rate \citep{matteucci2001,
mannucci2006} and the lifetime of a low-metallicity 5~M$_{\odot}$ star
\citep[e.g.,][]{schaller1992, karakas2007} are comparable
($\sim 80-100$ Myr), this apparently peculiar enrichment
scenario seems reasonable. \citet{mannucci2006} proposed that half of
the SNe~Ia associated with a starburst should explode on a timescale
of the order $\sim 100$ Myr. Marcolini et~al. (2006,2008) have shown
that even if a timescale greater than 1--2 Gyr is needed to appreciate
the cumulative effect of SNe~Ia in the mean [$\alpha$/Fe] ratio of
spheroidal galaxies, the effect of a single SN~Ia 
inhomogeneously polluting the ISM could be important in the first few
hundred Myrs \citep[see also][for and application of this scenario to the 
particular case of $\omega$~Cen]{marcolini2007}.

\section{Two-Region Model}
\label{sec:model}

\begin{figure}    
\begin{center}    
\psfig{figure=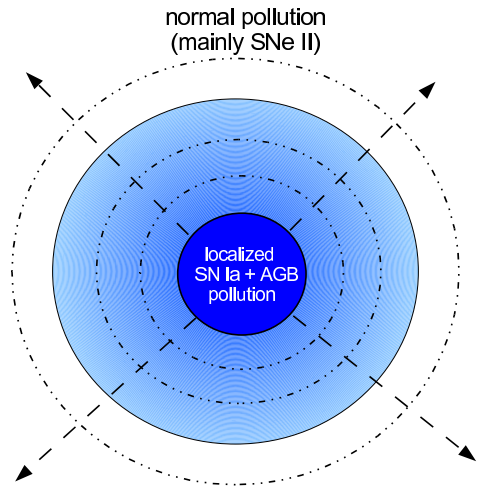,width=0.35\textwidth}
\end{center}   
\caption{Schematic of the model. Initially, a localised volume (inner
blue region) arises with the pollution from a single SN~Ia and
multiple intermediate-mass AGB stars.  After a new generation of stars
is born the associated SNe~II begin to pollute and expand the inner
volume, as well as mix it with the surrounding lower-[Fe/H]
interstellar medium.}
\label{fig:model} 
\end{figure}

Our model is based on what could be termed a ``two-region'' chemical
evolution framework with a simplified prescription for the chemical
enrichment.  As a first attempt to reproduce the main chemical
properties of GC stars, we do not follow the ejecta of different mass
progenitors, but we use IMF-averaged yields for both
intermediate-mass AGB stars and SNe ejecta (see~\ref{sec:yields} for
more detail). A sketch of the model is shown in Fig.~\ref{fig:model},
where the two regions -- ``inner'' and ``external'' -- are shown. The
dark blue ``inner region'' in Fig.~\ref{fig:model} corresponds
initially to the blue-green circle in the last panel of
Fig.~\ref{fig:model_initial} polluted inhomogeneously by a SN~Ia and
intermediate-mass AGB material; we will assume that the SF related to
the GC formation only takes place inside this region. In the
following, we will use the term ``inhomogeneous pollution'' to denote
that this region has different chemical properties from the
surrounding medium (note that the chemical properties inside the inner
region itself are homogeneous).

The external region is simply described by the initial value of
[Fe/H]$_{\rm ISM}$ given in Table~\ref{tab:gc_mod}, which should
reflect the mean chemical properties of the proto-halo gas at the
epoch of GC formation (enriched mainly by SNe II).  The inner region
is initially described by its radius (R$_{\rm in}$ i.e., the degree to
which the SN~Ia was initially confined) and the mean chemical
properties within, which is set by the number of 4$-$7~M$_{\odot}$ AGB
stars, $N_{\rm AGB}$. Since most of the iron in the inner region is
provided by a single SN~Ia the value of R$_{\rm in}$ also defines the
value of [Fe/H]$_{\rm in}$ in the inner region. It is obvious that the
localized iron-rich ejecta from the SN~Ia will substantially increase
the value of [Fe/H] inside the inner region compared to the value in
the external region, whereas intermediate-mass AGB stars increase the
N, Na, the neutron-rich Mg isotopes and, possibly, Al. A
glance at Table~\ref{tab:gc_mod} shows that the inner region is a
factor of $\sim 10-100$ more [Fe/H] rich than the external ISM and
also has a similar higher overall metallicity. For this reason the
first generation of stars will born with peculiar chemical properties:
they will be depleted in O and Mg (because of the single SN~Ia) and
they will be enhanced in N, Na, and Al (due to AGB pollution). Note
that in canonical AGB self-pollution models \citep[e.g.][]{fenner2004}
these stars are the last to form. While a detailed discussion about
each element will be presented in the following sections, here we
would like to point out that, in general, the trend for SN~Ia is to
decrease the [$X_{\rm i}$/Fe] value when $X_{\rm i}$ is not (or
negligibly) produced in intermediate-mass AGB stars (e.g., O and the
elemental Mg abundance to a lesser extent). In the case when the
elements are overproduced in AGB stars (e.g., N, Na and, possibly,
Al), the effect of the SN~Ia is to mitigate their enhancement.

\begin{table} 
\centering
\begin{minipage}{80mm} 
\caption{Initial conditions of the model. Note that
only three of the parameters are free as the value of [Fe/H]$_{\rm in}$ is 
defined by R$_{\rm in}$ (see text for more details).}
\label{tab:gc_mod}
\begin{tabular} {|l|c|c|c|c|c|} 
\hline 
 Model  & [Fe/H]$_{\rm ISM}$ & [Fe/H]$_{\rm in}$ & R$_{\rm in}$ (pc) & $N_{\rm AGB}$ \\ 
\hline
NGC~6752    &  -2.55     &  -1.60      &     36         &     250   \\
M~13        &  -3.50     &  -1.50      &     31         &     170   \\
NGC~2808    &  -2.85     &  -1.10      &     24         &     180   \\
\hline
\end{tabular}   
\end{minipage}   
\end{table}

We assume that the star formation takes place only inside the inner
region. As the SF proceeds, massive stars (M$_{\odot} \ge 8$
M$_{\odot}$) start exploding \citep[as early as $\sim 3-4$~Myr
after the onset of star formation ][]{schaller1992} as SNe~II,
ejecting freshly synthesized metals.  We assume that after each SN~II
explosion the inner region expands, driven by the explosions
themselves, mixing the inner-region gas with the surrounding ISM,
while the SF is still proceedings inside this region. During the
evolution, we simply assume that the density remains constant and thus
the mass inside a fixed radius is M= $\frac{4}{3} \pi \rho_{0} {\rm
R}^3$. For each of the proposed GC models we assume a density of
$\rho_{0}=4 \times 10^{-24}$ g cm$^{-3}$
which, for example, corresponds to a total mass of M$_{\rm ISM}\simeq
3 \times 10^{7}$ M$_{\odot}$ inside a 500~pc radius sphere. Even if we
were to assume a lower SF efficiency, there would still be more than 
sufficient gas to form a typical GC.

The simple relations governing the evolution after each SN~II
explosion can be summarized as:
\begin{equation}
R(t+\Delta t)=R(t) \times f_{\rm exp} 
\label{equation1}
\end{equation}
\begin{equation}
X_{i}(t+\Delta t)=X_{i}(t)+X_{i}(\Delta R)+\langle X_{i}(SN II)\rangle 
\label{equation2}
\end{equation}
\noindent
where $f_{\rm exp}$ is a constant governing the logarithmic expansion
of the inner region after the explosion of every single SN~II, $\Delta
t$ is the time interval between two successive SNe~II explosions,
$X_{i}$(t) is the mass fraction of an element $X$ at a time $t$ within the
radius R(t), $X_{i}$($\Delta$R) is the mass fraction of an element $X$
initially present in the corona R$(t+1)-$R$(t)$, and finally, $\langle
X_{i}(SN II) \rangle$ is the average mass fraction of element $X_i$
ejected by a single SN~II (see Sec.~\ref{sec:yields}). For example, the
evolution of the [Fe/H] abundance in the inner region at a time $(t+
\Delta t)$ (where $\Delta t$ is the time between two successive
SNe~II) is computed by adding the Fe initially present in this region
at R$(t)$ to the iron ejected by the newly exploding SNe~II and to the
iron initially present in the external region R$(t+\Delta
t)-$R$(t)$. This number is then divided by the total amount of
hydrogen inside the new radius R$(t+\Delta t)$.  All the other
variables are calculated accordingly. We assume that this
expansion is described by a logarithm law with a free parameter
(usually of the order of $f_{\rm exp}\sim$ 1.001) which fixes the
increase of R$(t)$ after each SN~II explosion (i.e., R$(t+\Delta t)=
$R$(t) \times f_{\rm exp}$).

\begin{table*} 
\centering 
\begin{minipage}{160mm} 
\caption{Mean SNe II stellar yields averaged over the progenitor mass
range 10--60 M$_{\odot}$ for a Salpeter IMF and for different authors:
W\&W=\citet{woosley1995}; C\&L=\citet{chieffi2004};
KOB=\citet{kobayashi2006}. The yields of Fe and He are given in solar
masses while for different elements we show the [$X_{\rm i}$/Fe]
ratios.}
\label{tab:sne_mod} 
\begin{tabular} {|l|c|c|c|c|c|c|c|c|c|c|c|c|c|} 
\hline
SNe II model &$\;\;\;$ Fe $\;\;\;$& $\;\;\;$ He $\;\;\;$& $\;\;$[C/Fe]$\;\;$&
$\;\;$[N/Fe]$\;\;$& $\;\;$[O/Fe]$\;\;$&$\;\;$[Na/Fe]$\;\;$&
$\;\;$[Mg/Fe]$\;\;$&$\;\;$[Al/Fe]$\;\;$ \\
\hline
W\&W (Z=0.0002 )  &  6.1e-2    & 6.7 & 0.13   & $-1.73$ & 0.51 & $-0.39$ &  0.15 & $-0.24$ \\
W\&W (Z=0.002)    &  6.9e-2    & 6.6 & 0.07   & $-0.86$ & 0.46 & $-0.16$ &  0.16 & $-0.04$ \\
C\&L (Z=0.0001)   &  1.0e-1    & 5.7 & 0.41   & $-1.91$ & 0.63 &  0.05   &  0.61 & $-0.14$ \\
C\&L (Z=0.001)    &  1.0e-1    & 5.7 & 0.40   & $-0.96$ & 0.63 &  0.26   &  0.61 &  0.11 \\
KOB  (Z=0.001)    &  7.5e-2    & 5.7 &$-0.16$ & $-0.80$ & 0.64 &  0.05   &  0.63 &  0.15 \\
KOB  (Z=0.001+HN) &  9.3e-2    & 5.7 &$-0.26$ & $-0.89$ & 0.52 & $-0.08$ &  0.52 &  0.28 \\
\hline 
Model (SNe II)    &  9.0e-2    & 6.0 & $-0.10$ & $-0.90$ & 0.50 & $-0.20$ &  0.50 & $-0.10$ \\
\hline 
\end{tabular}   
\end{minipage}  
\end{table*}

In Fig.~\ref{fig:model} the dashed lines represent a sketch of the
evolution of the inner region and the fact that the initial
contribution of the single SN Ia (as well as AGB stars) is diluted
during its expansion and the subsequent mixing with the outer, lower
metallicity ISM. As soon as the first SNe II start to explode there
are two contrasting mechanisms shaping the chemical properties of the
newly forming stars.  First, the freshly ejected SNe II metals
directly pollute the gas where new stars are forming. Second, the 
inner volume is expanding and the gas inside this
region is mixed with lower metallicity material.  In Sec.~\ref{sec:feh}
we will show that, during the chemical evolution process, it is
possible to keep the [Fe/H] abundance constant in the inner region
(where the new stars are forming) owing to these two
contrasting mechanisms.

While stars with peculiar properties form during the first stage,
during the second stage the chemical properties of the forming stars
$\alpha$-enhanced) as SNe II start exploding. 
\citet{marcolini2006} showed, for the
specific case of dwarf galaxies, that even if a single SN Ia can
temporarily pollute a small volume and dramatically change its
chemical composition, its imprint disappears almost completely once the
gas inside this volume mixes with the ISM. In the next section
we will show that the last stars forming in our model (stars with
[O/Fe]$\sim0.5$) do not show any signature of the former AGB or 
SN Ia pollution, having chemical properties typical of 
pure SNe II ejecta.

We will also show, in Sec.~\ref{sec:feh}, that in our model the inner
region is  more metal rich than the outer region and, accordingly,
the cooling is much more efficient there. Therefore, this region is
more likely to host an extremely enhanced SF -- typical of GCs --
than the outer ISM (where normal halo
stars are forming). 
 
Numerous authors have discussed, from an energetics perspective, the
possibility that stellar clusters can experience SNe~II self-enrichment, 
and there is a wide range of evidence as to whether, 
and to what extent, this is possible in GCs 
\citep[e.g.][]{dopita1986, morgan1989,
smith1996, parmentier2001, gnedin2002, thoul2002, parmentier2004,
recchi2005, prantzos2006}.  
For example, it has been suggested that stellar winds and SNe
explosions from the first generation of stars can form a
metal-enriched supershell, where further star formation is triggered
\citep{brown1991, brown1995, thoul2002, parmentier2004,
recchi2005}. \citet{brown1991} concluded that a small
entirely self-enriched system of stars typically achieves a metallicity
in the range $10^{-2}--10^{-1}$ Z$_{\odot}$, and that the second generation
of stars would be expected to be extremely homogeneous in composition.
\citet{melioli2004} studied in detail the heating efficiency
(i.e., the fraction of SNe energy which is not radiated away) of SNe
exploding in a starburst environment with properties similar to that
of our model. These authors found that the heating efficiency can be
very close to zero (i.e., all the SNe energy is efficiently radiated
away) for, up to, a few tens of Myr, before rapidly increasing to unity and
leading to a possible galactic wind. While we cannot follow the
dynamical evolution (and the energetics) of our model, it is possible
to speculate that this second stage of star formation could be as rapid
as $\sim 20-40$~Myr (with SNe II directly polluting the newly forming
stars), or it could proceed in a long series of weak bursts if the
star formation is self-regulated - i.e., the gas is temporarily expelled
out of the star forming regions by the SNe and re-collapses (due to the
outer pressure) once the SNe~II energy is radiated away. If the star
formation is self-regulated, the SF history can be as long as
$\sim$100~Myr. A longer SF history would require the contribution of
AGB and SNe~Ia self-pollution, which does not seem to be required in our
model, to explain the chemical properties of GCs. Finally, in our case,
the formation of a galactic wind would be associated with the
cessation of the SF itself and it would happen after $\sim 1-5 \times
10^{3}$ SNe~II explosions, depending upon the details of the
model.  Although these earlier studies seems to support the formation
of second generation of stars after the occurence of an explosive event, 
and the formation of a bound system, 
hydrodynamical simulations of the SNe~II self-enrichment
phase in quantitative detail are needed to further quantify the 
star-formation timescales and the viability 
of our model from a dynamical point of view.
These will be presented in a forthcoming paper in this series.

\begin{table*} 
\centering 
\begin{minipage}{165mm} 
\caption{Mean SNe II, SNe Ia and intermediate-mass AGB yields in solar
masses. SNe II yields refer to the model in
Table~\ref{tab:sne_mod}. SNe Ia yields refer to the WDD1 model of
\citet{iwamoto1999}. The mean AGB yields are averaged over the
progenitor mass range 4--7 M$_{\odot}$ using a Salpeter IMF for
different authors: KA07=\citet{karakas2007}; FE04=\citet{fenner2004};
KA03=\citet{karakas2003}; IZ07=\citet{izzard2007}; all the yields are
for $Z=0.0001$ except for the case of \citet{fenner2004} which is at
[Fe/H]=$-1.40$. The yields of Fe and He are given in solar masses.}
\label{tab:agb_mod} 
\begin{tabular} {|l|c|c|c|c|c|c|c|c|c|}
\hline  
Polluter        &$\;\;\;\;\;$ Fe $\;\;\;\;\;$&$\;\;\;\;\;$ He $\;\;\;\;\;$&
$\;\;\;\;\;\;$ C $\;\;\;\;\;$ &$\;\;\;\;\;$ N $\;\;\;\;\;$&$\;\;\;\;\;$ O $\;\;\;\;\;$&
$\;\;\;\;\;$ Na $\;\;\;\;\;$ &$\;\;\;\;\;\;$ Mg $\;\;\;\;\;$&$\;\;\;\;\;$ Al $\;\;\;\;\;$ \\
\hline
Model (SNe II)  & 9.00e-2 & 6.00    & 1.59e-1 &  7.66e-3 &  1.82  & 1.60e-3 & 1.49e-1 & 3.23e-3 \\
\hline
Model (SNe Ia)  & 6.72e-1 & 5.66e-3$^a$ & 5.42e-3 & 2.85e-4 & 8.82e-2 & 8.77e-5 & 7.69e-3 & 4.38e-4 \\
\hline
AGB (KA07)  & 2.27e-5 & 1.45   & 6.21e-3 & 6.89e-2 & 1.44e-3 & 1.60e-3 & 2.17e-3 & 7.01e-5 \\
AGB (FE04)  & $\times$&$\times$& 8.54e-3 & 5.76e-2 & 1.86e-3 & 5.95e-4 & 1.52e-3 & 6.13e-5 \\
AGB (KA03)  & $\times$&$\times$& $\times$& $\times$& $\times$& 7.79e-5 & 1.70e-3 & 1.04e-4 \\
AGB (IZ07)  & $\times$&$\times$& $\times$& $\times$& $\times$& 2.92e-5 & 2.15e-3 & 2.02e-4 \\
\hline
Model (AGB) &  2.27e-5 & 1.45  & 1.55e-3 & 6.89e-2 & 1.44e-3 & 4.00e-4 & 2.17e-3 & 3.50e-3 \\
\hline 
\end{tabular} 
$^a$ the value of He for SNe Ia is from \citet{kobayashi2006}.
\end{minipage} 
\end{table*}

\section{Yields}
\label{sec:yields}

As noted earlier we used IMF-averaged yields for both SNe~II and 
intermediate AGB polluters.
For SNe~II, we explore the effect of using yields computed by
different authors \citep[][see
Table~\ref{tab:sne_mod}]{woosley1995,chieffi2004, kobayashi2006} and
use the limits 10--60 M$_{\odot}$.  Previous studies have found that
the \citet{woosley1995} models tend to overproduce the amount of Fe --
with a subsequent underestimation of the [$\alpha$/Fe] abundances --
for this reason we half Fe values as done in several studies \citep[e.g.,][]{timmes1995,
goswami2000, fenner2004}.  While the set of yields
are in reasonable agreement with one another for most of the elements,
occasional differences as large as 0.4~dex are present
\citep[e.g.,][]{gibson1997}. We computed a set of yields that we
use throughout the paper unless stated otherwise. These yields are the
average value of the above mentioned theoretical set of yields when
the agreement between them is good (when they differ less than
0.4~dex). Otherwise, we chose the yields that best reproduce the
low metallicity ($-2.0 \le$[Fe/H]$\le -1.0$) halo stars
\citep[e.g][]{gratton2000}.  These computed values are labelled as
``Model'' yields in Table~\ref{tab:sne_mod}.

For SNe~Ia we use the yields of \citet{iwamoto1999}, specifically
their case WDD1 (note however that for the elements considered in this
paper a different choice of their SNe~Ia model would not affect the
results). For a consistent chemical evolution model, the 
metallicity of the pre-polluting SN Ia and AGB stars
should be be roughly the same as the external region
([Fe/H]$_{\rm ISM}$), while the metallicity of the self-polluting
SNe~II match that of the inner region ([Fe/H]$_{\rm
in}$). Unfortunately, the set of yields of \citet{iwamoto1999} are
only available for solar metallicity and we are not aware of 
low-metallicity SNe Ia yields available in the literature.

In Table~\ref{tab:agb_mod} we summarize the mean yields for
intermediate-mass AGB stars, calculated by averaging the $Z=0.0001$
values from \citet{karakas2007} using a Salpeter (1955) IMF in the mass
range 4--7 M$_{\odot}$ (we will refer to this model as the ``reference
model''). As a comparison we also show the yield values obtained averaging
similar datasets at the same metallicity published by
\citet{karakas2003} and \citet{izzard2007}, as well as the yields
calculated by \citet{fenner2004} for the specific case of AGB
self-pollution in NGC~6752 at [Fe/H]$=-1.40$. In general, each set is
in good agreement although there are significant differences present
for some elements (e.g., Na yields span $\sim 2$ orders of
magnitude). A detailed study of how the production of the Ne, Na, Mg,
and Al isotopes in AGB stars are effected by reaction rate
uncertainties was carried out by \citet{izzard2007}. They concluded
that the most uncertain yields are those of $^{26}$Al and $^{23}$Na,
with variations of two orders of magnitude. The yields of
$^{24}$Mg and $^{27}$Al have typical uncertainties of one order of
magnitude. These uncertainties are comparable to those resulting from
stellar modelling uncertainties, where differences of one order of
magnitude or more can be found by varying the mass-loss rate or
convective model \citet{ventura2005a,ventura2005b} and
\citet{karakas2006c}.

For this reason we also decided to use a set of $ad-hoc$
intermediate-mass AGB yields that represent the different element
production that we would require in order to reproduce the
observational constraints in the framework of our model.  This set of
yields is presented in Table~\ref{tab:agb_mod}
as ``Model''. These values are within the limits of the
above-mentioned set of yields, and always within a factor of four of
the values of the reference model (with the exception of Al see
Section~\ref{sec:almg}). In the following, the models using
\citet{karakas2007} (reference model) will be shown as solid lines,
while the models using our proposed set of yields will be shown
as dashed lines.

We would like to stress that the AGB yields we propose in
Table~\ref{tab:agb_mod} should not be taken as absolute values but as
relative ratios between the different elements. This is because it is possible
to re-scale the yields by changing the number of $N_{\rm AGB}$ stars needed
to fit the observations. We use nitrogen as a reference element and
re-scale the other elements accordingly.  We choose nitrogen because
it is better constrained than the other elements (e.g., Na, Al) since
its production does not depend as strongly on reaction rate
uncertainties. Note that the recent revision of the
$^{14}$N$(p,\gamma)^{15}$O reaction rate \citep{bemmerer06}, $\sim
60$\% slower at temperatures below 100$\times 10^{6}$K, would result
in even more N production.  Nitrogen yields are, however, effected by
other modelling uncertainties such as convection, mass loss and the
modelling of the third dredge-up \citep{marigo03,karakas08}.  For
example, an increase in the mass-loss rate would result in smaller N
yields because of a shortened HBB phase (and also less Na, Mg and Al).

\begin{figure*}    
\begin{center}    
\psfig{figure=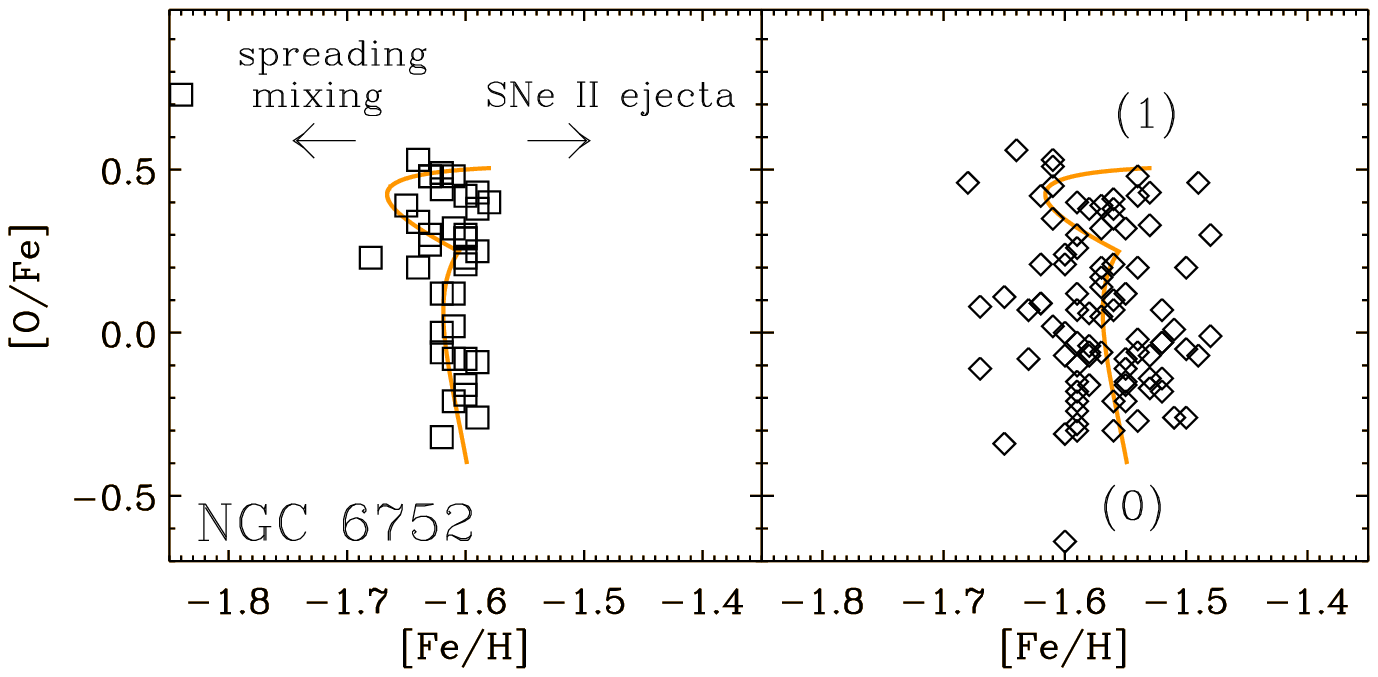,width=0.80\textwidth} 
\end{center}   
\caption{The predicted trend of [Fe/H] versus [O/Fe] compared with the
observational dataset of \citet{yong2005} (left panel) and
\citet{carretta2007} (right panel) for the case of NGC~6752. It is
possible to maintain an invariant [Fe/H] abundance owing to a tuning
between the two opposing mechanisms: SNe~II eject freshly synthesized
Fe (and O), which is mixed with gas from the inner (inhomogeneous)
volume along with material from the external ISM, which has a lower
[Fe/H] value.}
\label{fig:feo} 
\end{figure*}

\begin{figure}    
\begin{center}    
\psfig{figure=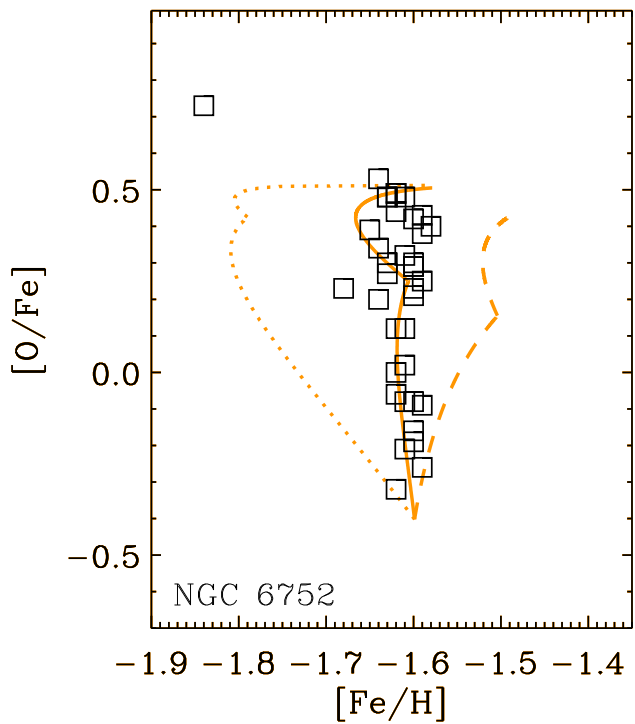,width=0.45\textwidth} 
\end{center}   
\caption{The predicted trend of [Fe/H] versus [O/Fe] for three models
with different $f_{\rm exp}$ values compared with the observational
dataset of \citet{yong2005} for the case of NGC~6752. The initial
values of $f_{\rm exp}$ are 1.0003 (dashed line), 1.0009 (solid line)
and 1.0027 (dotted line) and are increased to 1.0012, 1.0035, and 1.015
after 750 SNe II have been exploded. The models are stopped after
$\sim$1600 SNe II have been exploded polluting the ISM. It is
evident that changing the value of $f_{\rm exp}$ within a factor of three
effects the [Fe/H] ratio of the forming stars by only 0.1-0.2 dex.}
\label{fig:feo} 
\end{figure}

\begin{figure*}    
\begin{center}    
\psfig{figure=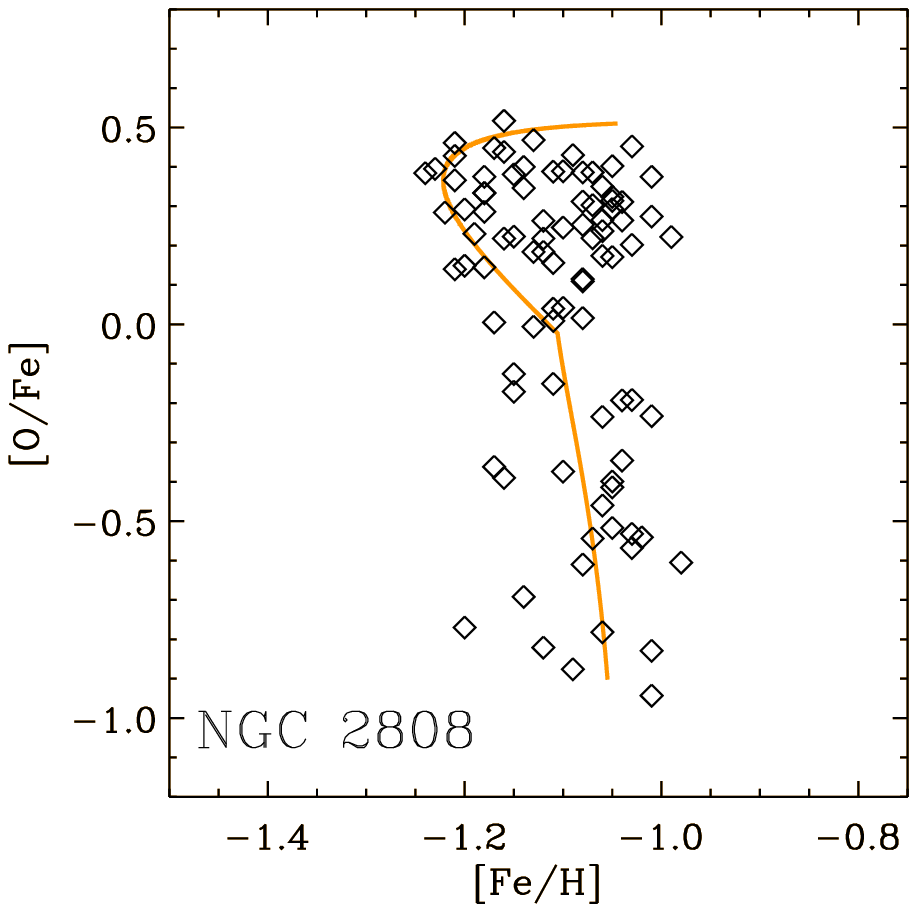,width=0.40\textwidth} 
\psfig{figure=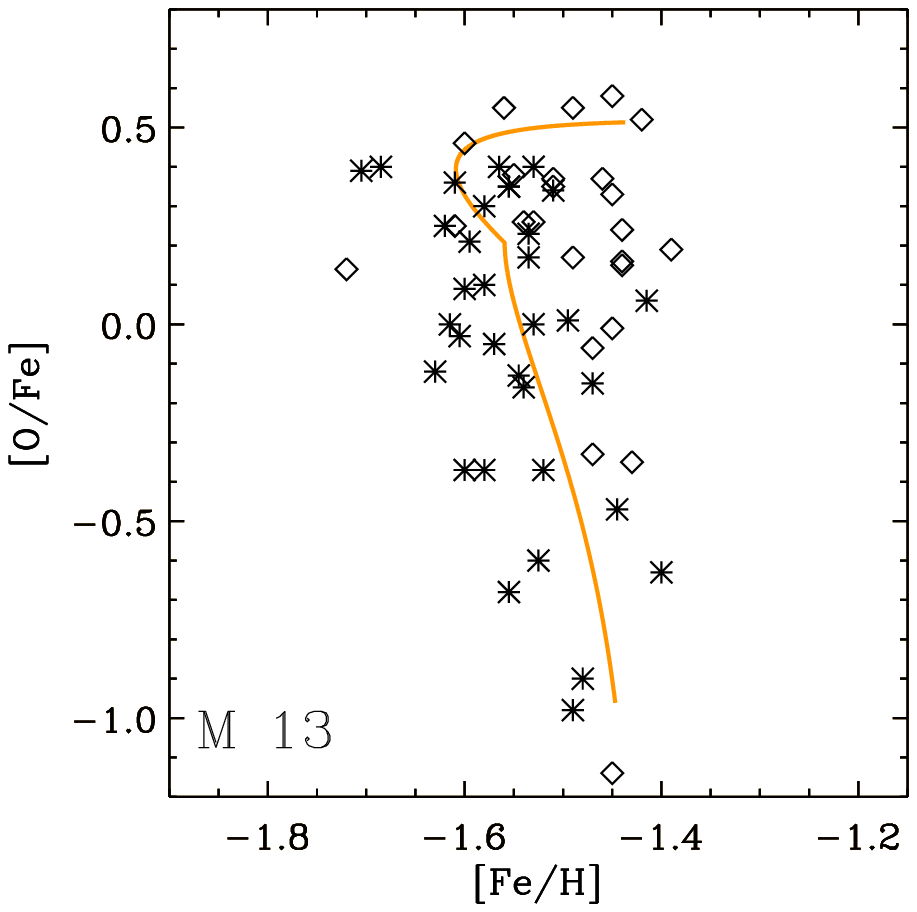,width=0.40\textwidth} 
\end{center}   
\caption{The predicted trend of of [Fe/H] versus [O/Fe], 
for the case of NGC~2808 (left panel), compared with the
observational dataset of \citet{carretta2006}. The case of M~13
(right panel) is plotted against the observational dataset of
\citet[][asterisks]{sneden2004} and \citet[][diamonds]{cohen2005}.}
\label{fig:feo_bis} 
\end{figure*}

\begin{figure*}     
\begin{center}    
\psfig{figure=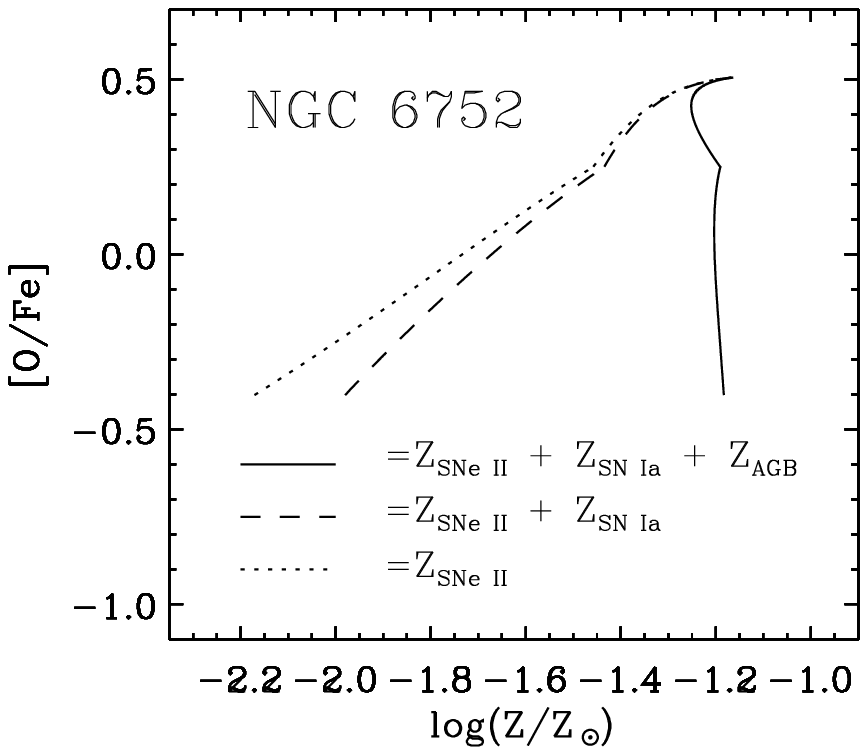,width=0.32\textwidth}
\psfig{figure=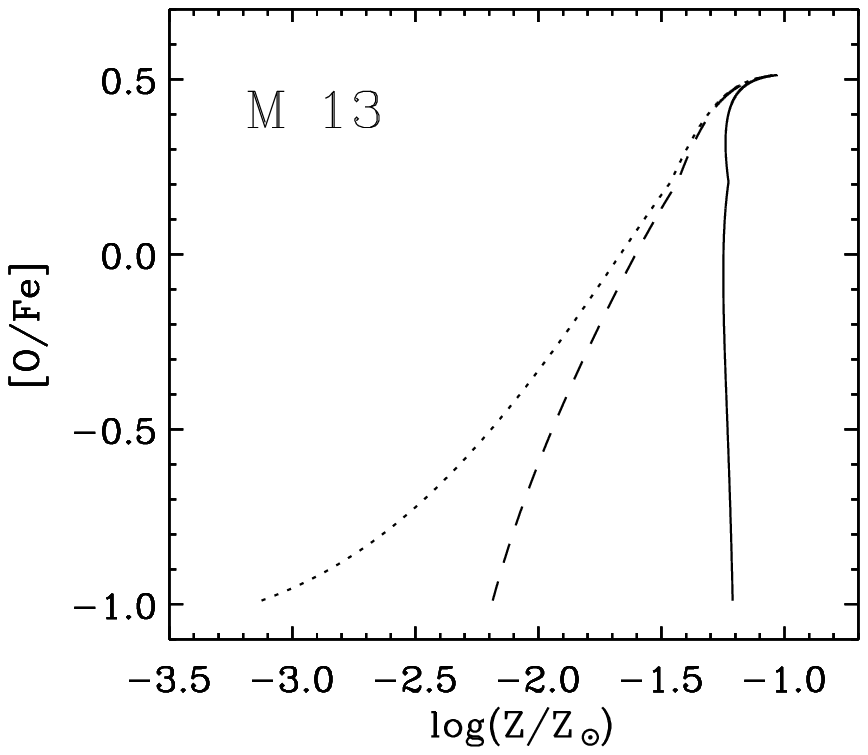,width=0.32\textwidth}
\psfig{figure=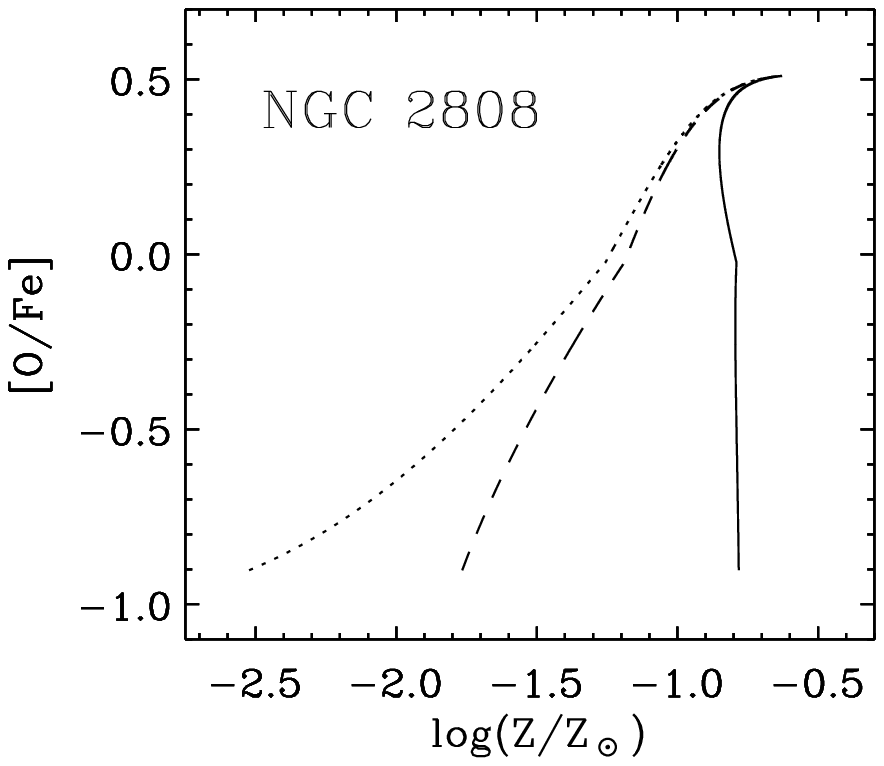,width=0.32\textwidth}
\end{center}   
\caption{Evolution of the metallicity $\log(Z/Z_{\odot})$ versus
[O/Fe] in the inner region for the three reference models: NGC~6752
(left panel), M~13 (central panel), and NGC~2808 (right panel). The
dotted line corresponds to the case where we ignored the contribution
of the SN~Ia+AGB pollution $only$ to the total metallicity
$\log(Z/Z_{\odot})$ (note that low initial [O/Fe] value can be
achieved only due to the localized contribution of the SN Ia), while
dashed lines correspond to the model with no pollution from AGB
stars. Note that the initial value of $\log(Z/Z_{\odot})$ (thus at low
values of [O/Fe]) for the dotted line corresponds to the value of the
metallicity for the external region. Most of the
metals in the inner region are initially produced by AGB stars while at
the end of the evolution only SNe~II contribute to the metal content.}
\label{fig:zstelle} 
\end{figure*}

\section{Results: the case of NGC~6752, M~13 and NGC~2808}
\label{sec:gc_discussion}

To test our framework in detail, we will focus on three 
well studied GCs: NGC~6752, M~13, and NGC~2808. All these GCs have
an intermediate metallicity ranging from $-1.60 \le$[Fe/H]$\le -1.10$
\citep[e.g.,][]{harris1996}. While NGC~6752 can be defined as a
``normal'' GC, NGC~2808 and M~13 show extreme values of the
[O/Fe] ratio (down to $\sim$-1.0) and are excellent cases to test our
new proposed framework. In this section we show how different
choices of the initial conditions in the pre-enrichment during the MW
halo formation can led to different chemical peculiarities in the
forming GC.

\subsection{[Fe/H] and Metal content}
\label{sec:feh}

Before going any further, it is important to follow the iron content
of the forming stars in our model. As previously
mentioned, the possibility of SNe II and SN Ia  self-enrichment in GC have
not been considered in the literature before,  
because they produce large amounts of 
Fe, whereas the [Fe/H] content in most of the GCs appears to be constant.

The initial conditions of NGC~6752 are summarized in
Table~\ref{tab:gc_mod}: the R$_{\rm in}$ of the inhomogeneous region
polluted by the SN Ia is 36 pc and this volume is also polluted by 250
intermediate 4--7 M$_{\odot}$ AGB stars. The mean iron content of the
external region is [Fe/H]$=-2.55$, reflecting the early stage of the
halo formation, while the effect of the SN Ia increases the value of
[Fe/H] up to $-1.60$ inside the inner region. In the same table, we
also summarize the initial parameters for the cases of M~13 and
NGC~2808.
    
For each GC, the initial conditions are constrained by the
observed stellar values of [Fe/H] and the peculiar values of
[O/Fe]$_{\rm min}$ and [Na/Fe]$_{\rm max}$ (or alternatively the value
of [N/Fe]$_{\rm max}$), as each of these values depends mainly on one
polluter.  Since most of the iron in the inner region is
produced by SN Ia, the initial [Fe/H]$_{\rm in}$ content of the
forming stars is set by the value of R$_{\rm in}$, while oxygen is
produced mainly by SNe II and Na by intermediate-mass AGB stars (see
Tables~\ref{tab:sne_mod} and \ref{tab:agb_mod}). We assume that a SN Ia
always deposits the same amount of Fe (0.74 M$_{\odot}$), hence decreasing
(increasing) the value of R$_{\rm in}$ has the effect of decreasing
(increasing) the hydrogen content inside the inner region and 
increasing (decreasesing) the [Fe/H] content (this is evident in
Table~\ref{tab:gc_mod}, where the value of [Fe/H]$_{\rm in}$ 
increases for lower values of R$_{\rm in}$). Once the [Fe/H] content in
the inner region is set, the overall [Fe/H]$_{\rm ISM}$ value
constrains the initial [O/Fe]$_{\rm min}$ of the model: this is because,
according to a pure SNe II pollution model, the O content is 
proportional to the Fe content (and the Fe produced by SNe II can 
be ignored in the inner region). 
At this point, the observational value of [Na/Fe]$_{\rm max}$ is 
sufficient to constrain the number of intermediate-mass AGB stars 
needed to reproduce the Na-O anti-correlations.

For example NGC~6752 and M~13 have similar initial values of
[Fe/H], at $\simeq-1.60$ and [Fe/H]$\simeq-1.50$, respectively, that
reflects comparable values of R$_{\rm in}$. In contrast the
``external'' values of [Fe/H]$_{\rm ISM}$ for each cluster differ by
an order of magnitude (see Table~\ref{tab:gc_mod} reflecting, as shown
later, their different extreme values of [O/Fe]. In contrast,
NGC~2808 shows an even higher value of [Fe/H]$\simeq -1.10$ because of
its smaller value for R$_{\rm in}$=24 pc, even if its [Fe/H]$_{\rm
ISM}$ is slightly lower than for the case of NGC~6752 (see
Table~\ref{tab:gc_mod}).  

In Fig.~\ref{fig:feo} we show the [Fe/H] versus [O/Fe] abundance
evolution of the reference model for NGC~6752. Data from \citet{yong2005}
and \citet{carretta2007} is included for comparison.
To compare our model with the 
Carretta's data, we apply an offset of 0.1~dex to our predicted value 
of [Fe/H]. This is  to take into account a similar difference in the mean 
[Fe/H] values between the two datasets. It is
clear from the figure that the [Fe/H] content remains relatively
constant during the whole evolution and it is consistent with the
observational data. Likewise, the evolution of [O/Fe] increases
monotonically from its initial peculiar value [O/Fe]$\simeq-0.4$ 
to 0.5.

The [Fe/H] abundance stays constant during the evolution because of
the tuning between the two mechanisms changing the Fe content of the
forming stars. While the newly exploding SNe II eject freshly
synthesized Fe, increasing the [Fe/H] content, their action also has
the effect of expanding the inhomogeneous inner region, mixing the
inner material with the external ISM (with a lower [Fe/H] content). 
Both these two contrasting effects are controlled, in our model, 
by the parameter $f_{\rm exp}$ (i.e., the rate of expansion of
R$_{in}$ after each SN II explosion, see Eq.~1) which increases
logarithmically with time.  It is evident from Fig.~\ref{fig:feo}
that, for this particular model, the effect of the
expansion is, at  the beggining, lowering  the [Fe/H] content, whereas 
at a later stage the SNe II
pollution becomes increasingly dominant. The [O/Fe] 
abundance of the forming stars starts  increase monotonically during 
the evolution, as the Fe content is practically constant and 
SNe II ejects a large amount of O (a mean SN II produces
1.82 M$_{\odot}$ of oxygen). Once the inhomogeneous region starts
expanding, the O content of the forming stars reaches the asymptotic
value of [O/Fe]=0.5, typical of pure SNe II pollution.

\begin{figure*}    
\begin{center}    
\psfig{figure=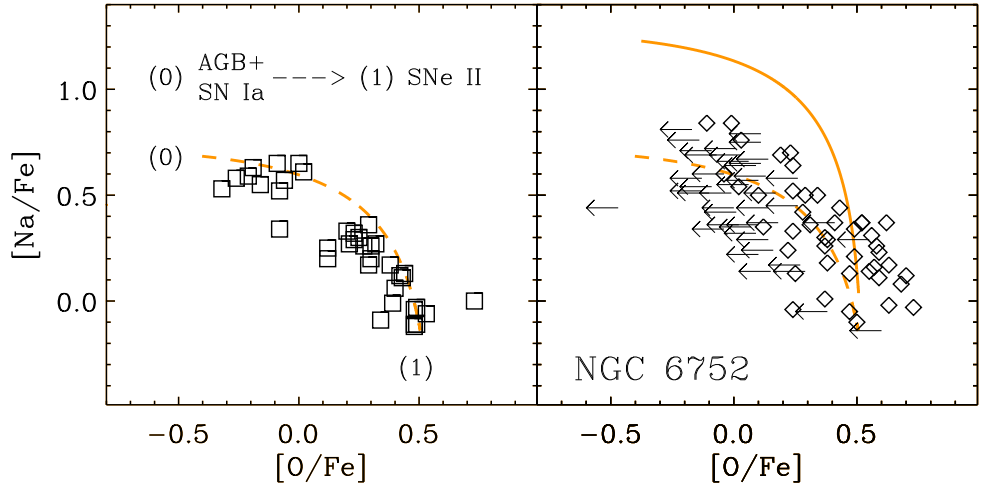,width=0.75\textwidth}
\end{center}   
\caption{The predicted evolution of the [Na/Fe] versus [O/Fe] abundance 
for NGC~6752, plotted against the observational datasets of
\citet{yong2005} (left panel) and \citet{carretta2007} (right panel,
arrows refer to upper limit values; see the original paper for
details). In our framework the AGB+SN Ia pollution is due to
pre-existing halo stars while only SNe II self-pollute the GC.  The
solid line in the right panel refers to the reference model using the
yields of \citet{karakas2007} while the dashed lines refer to a model
with Na production lowered by a factor of four compared with the
reference model (see Table~\ref{tab:agb_mod}).}
\label{fig:nao} 
\end{figure*}

\begin{figure*}    
\begin{center}    
\psfig{figure=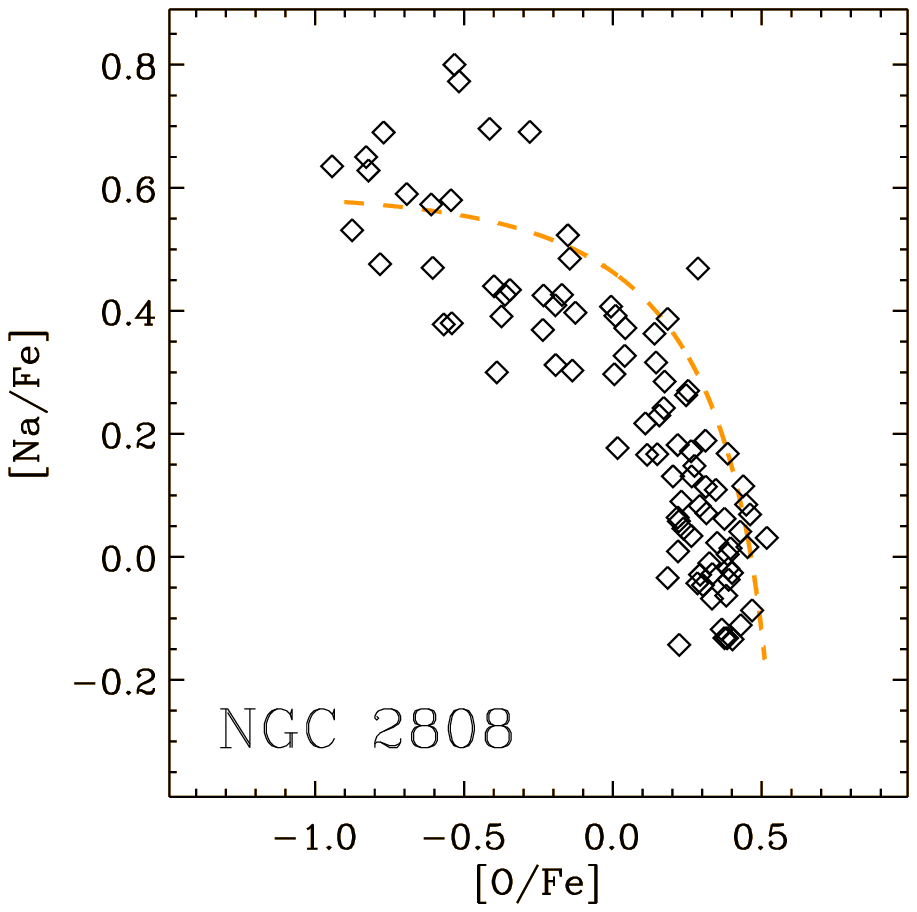,width=0.40\textwidth}    
\psfig{figure=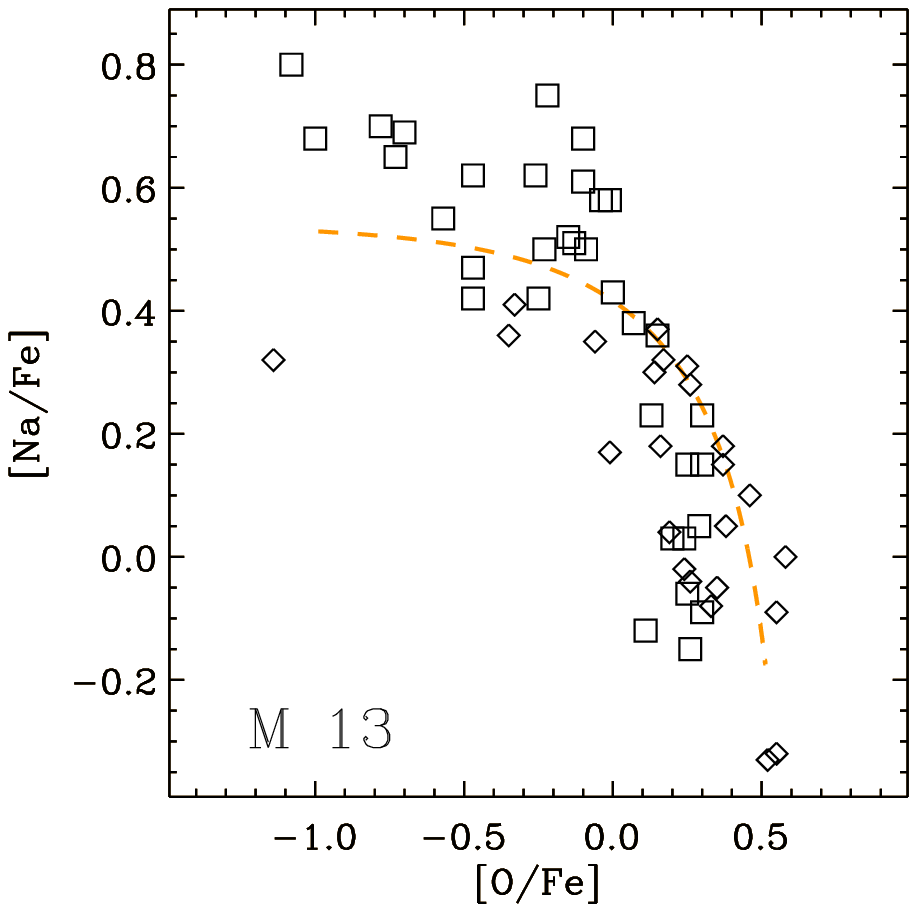,width=0.40\textwidth}
\end{center}   
\caption{(Left panel) the predicted evolution of the [Na/Fe] versus [O/Fe]
abundance for NGC~2808 against the observational data from \citet{carretta2006}.
(right panel) the same trend for the case of M~13 plotted against the
observational data from \citet{sneden2004} (squares) and
\citet{cohen2005} (diamonds).}
\label{fig:nao_bis} 
\end{figure*}

We use two different values of $f_{\rm exp}$ at two
different stages, as this gives a  better fit to the data.
For example, for NGC~6752, the initial value of 
1.0003 is increased to 1.001 after $\sim 10^3$ SNe II have 
exploded.  The effect on the model of this variation 
can be seen in Fig.~\ref{fig:feo} as an  small $"hook"$  at [O/Fe]$\sim$0.25.
Even if it were possible to better tune the model (by using 
another expansion law or simply by finding  the best parameter combination 
to maintain [Fe/H] perfectly constant during the evolution), we
are confident that such a fine tuning is not necessary and would
not significantly change our conclusions. Indeed, given all the 
other model uncertainties we think it is unnecessary to attempt to
fine tune the expansion caused by the SN II explosions.  However, to 
check the robustness of our conclusions to the assumed expansion value, 
we show in
Fig.~\ref{fig:feo} the predicted trend of [Fe/H] vs. [O/Fe]
for three different values of $f_{\rm exp}$
(roughly spanning a factor of 10 in dynamic range). 
It can be seen that changing
$f_{\rm exp}$ by a factor of three affects the [Fe/H] ratio of the
forming stars by only 0.1-0.2 dex.

In Fig.~\ref{fig:feo_bis} we compare the results from our model
with NGC~2808 and M~13. These GCs were chosen on the basis of their 
peculiar O depletion with values of [O/Fe] as low as $-1.0$. 
These low values are very difficult to obtain with only primordial
self-enrichment, which may explain the oxygen depletion down to a
minimum value [O/Fe]$\sim-0.5$ but no lower. Extra mixing may also
contribute to the depletion of oxygen but only for stars near the
tip of the first giant branch.
\citet{dantona2007} invoke a combination of the two to explain
the case of M~13, while in the model of \citet{bekki2007} such extreme
values are achieved by assuming that no third dredge-up occurs in 
 intermediate-mass AGB stars.

We are able to reproduce the low values of [O/Fe] without any
particular assumption (see Fig.~\ref{fig:feo_bis}). Indeed, such low 
[O/Fe] values arise naturally if the inhomogeneous region --
polluted by the initial SN Ia -- is smaller in the case of 
NGC~2808 and M13 than NGC~6752  (see Table~\ref{tab:gc_mod}).
In the case of M~13, the extreme low values of [O/Fe] of $-1.0$ 
requires a slightly smaller R$_{\rm in}$ as well as less oxygen from
the SNe II that pre-polluted the halo. To fit the
properties of M~13, the metallicity of the original ISM gas at 
the time of GC formation needs to be very low, with a typical [
Fe/H] abundance an order of magnitude lower ([Fe/H]$_{\rm ISM}=-3.50$) 
than for NGC~2808 or NGC~6752. This is despite 
the final [Fe/H] of M~13 is similar to NGC 6752.
The lower initial of [Fe/H]$_{\rm ISM}$ for M~13 than for NGC~6752 could indicate that this
GC formed at an earlier epoch and that it is older
than the other two GCs. Photometric ages of M~M13 compared to 
NGC~6752 suggest that M~13 is older \citep{rakos2005}, but 
these age determinations are highly uncertain and should
be treated as tentative.
 
In Fig.~\ref{fig:zstelle} we show the evolution of the logarithm of
the metallicity ($\log(Z/Z_{\odot}$) versus the [O/Fe] abundance
inside the inner region for the above models (solid
lines). As in the case of the [Fe/H] abundance, the total metallicity 
remains approximately constant during the evolution.  The dashed lines
represent the same evolution without taking into account the
contribution of metals by intermediate-mass AGB stars to the total
metallicity fraction. The dotted lines represent the case in which
the contribution of SN Ia and AGB to the total metallicity are ignored
(that is, only SNe II contribute to the global metallicity).
Most of the initial metals (e.g., C, N) inside the inner region 
are due to localized AGB pollution, while the SN Ia contribution 
is minimal. Indeed, even if a SNe Ia ejects large amounts of Fe 
(0.74 M$_{\odot}$) they eject only 1.4 M$_{\odot}$ of total metals.
Most of the ``metal'' are instead produced by AGB stars (an 
intermediate-mass AGB star produces a mean value $\sim$0.08 M$_{\odot}$ 
of metals, most of which is N; see Table~\ref{tab:agb_mod}). 
At the end of the evolution, most of the metals are produced by 
SNe II, where all lines converge to the same value in 
Fig.~\ref{fig:zstelle}. This is also demonstrated by the asymptotic
value of [O/Fe]=0.5 typical of pure SNe II pollution. Since the
metallicity of the stars is roughly constant during the evolution
this is also the metallicity of the forming stars.

We do not have a parameter that sets the fraction of SNe II that pollute 
the forming stars. Instead we calculate that assuming the initial mass 
of the GC is the same as it is today \citep{pryor1993}, and assuming 
a constant density and metallicity for the gas in the inner region, 
the fraction of SNe II ejecta that directly enriches the inner 30 
(50)~pc (i.e., where the GC star should be forming) turns out to be 
very low, of the order of 0.5-1\% (3-6\%), depending on the details 
of the model. Indeed, most of the metals produced by the SNe II are 
expelled beyond this limit of $\sim 50$~pc.

\subsection{The O-Na anti-correlation}
\label{sec:nao}

Intermediate-mass AGB stars do not produce Fe and O
\citep{karakas2007}, hence they have a negligible contribution to the
evolution of these elements. AGB stars do, however, produce a
considerable amounts of Na.

In Fig.~\ref{fig:nao} we show the evolution of [O/Fe] versus 
[Na/Fe] abundances predicted by our model, together with the observational 
data from \citet{yong2005} and \citet[][; upper limits are shown as
arrows]{carretta2007} for the case of NGC~6752. As already mentioned,
the first stars to form in our model are [Na/Fe]-rich (due to the
Na-rich AGB pollution) and [O/Fe]-depleted (due to the localized
iron-rich SN Ia ejecta). The effect of the SN Ia is to mitigate the
increase of the [Na/Fe] abundance that would otherwise reach much higher
values. For example, in Fenner et al. (2004) the [Na/Fe] abundance increased
to [Na/Fe]$\simeq$1.7. Once the first SNe II starts to pollute the
ISM and the inhomogeneous region expands, the chemical properties
of the forming stars evolve toward "normal" [Na/Fe]-poor and [O/Fe]-rich. 
In the figure we show two different predictions using 1) 
our reference model yields, and 2) the same yields but with the Na
yields from \citet{karakas2007} reduced by a factor of four.  While 
the reference model shows a clear O-Na anti-correlation, qualitatively in agreement
with the observations, it overestimates the initial value of [Na/Fe]
by $\sim$0.4-0.6~dex. In comparison, the model with the reduced Na
yields reproduces better the observed anti-correlation.  Note
that the reduced value ($\langle$Na$\rangle$=$4.0 \times 10^{-4}$
M$_{\odot}$) is  in better agreement with the Na production reported
by \citet{fenner2004} ($\langle$Na$\rangle=5.9 \times 10^{-4}$
M$_{\odot}$ M$_{\odot}$), and by \citet{karakas2007} for more
metal-rich AGB stars ($\langle$Na$\rangle$=7.7$\times 10^{-4}$
M$_{\odot}$ for $Z=0.004$). The reduced Na yields are also well
 within model uncertainties (see Table~\ref{tab:agb_mod}). It is 
also possible to obtain a very good match to the absolute values of 
[Na/Fe] by simply reducing (by same factor of 4) the amount of AGB 
pollution compared to the reference model (see Table~\ref{tab:gc_mod}), 
because the number of AGB stars has little effect on O and Fe 
(see Table~\ref{tab:agb_mod}). However, as discussed in 
Sec.~\ref{sec:yields}, we use AGB nitrogen production as a 
reference (see \S~\ref{sec:cn}). Thus, in the following we will 
always assume a mean value of 
$\langle$Na$\rangle$=$4.0 \times 10^{-4}$ M$_{\odot}$ from 
intermediate-mass AGB stars.

The [O/Fe] versus [Na/Fe] abundance evolution for the models 
tailored for NGC~2808 and M~13 are shown in Fig.~\ref{fig:nao_bis}, 
along with their respective observational datasets. The
O-Na anti-correlation is well reproduced for these two GCs, including
the extremely low values of [O/Fe].

\begin{figure}    
\begin{center}    
\psfig{figure=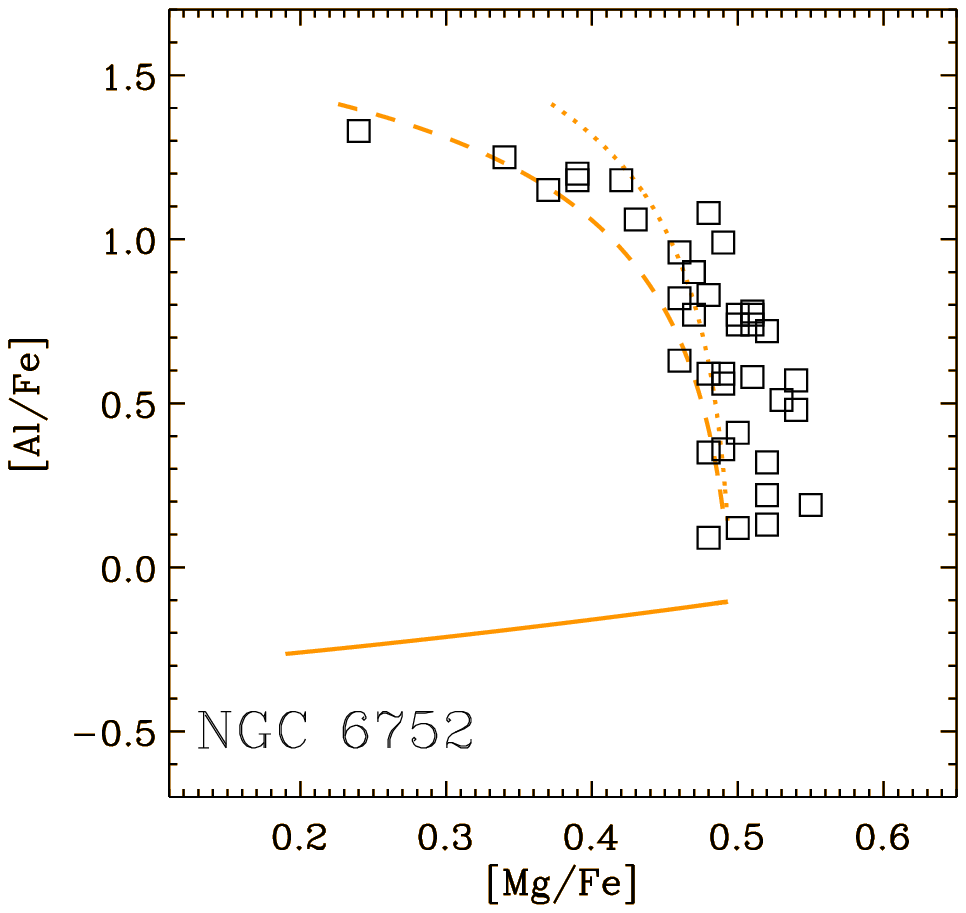 ,width=0.40\textwidth}
\psfig{figure=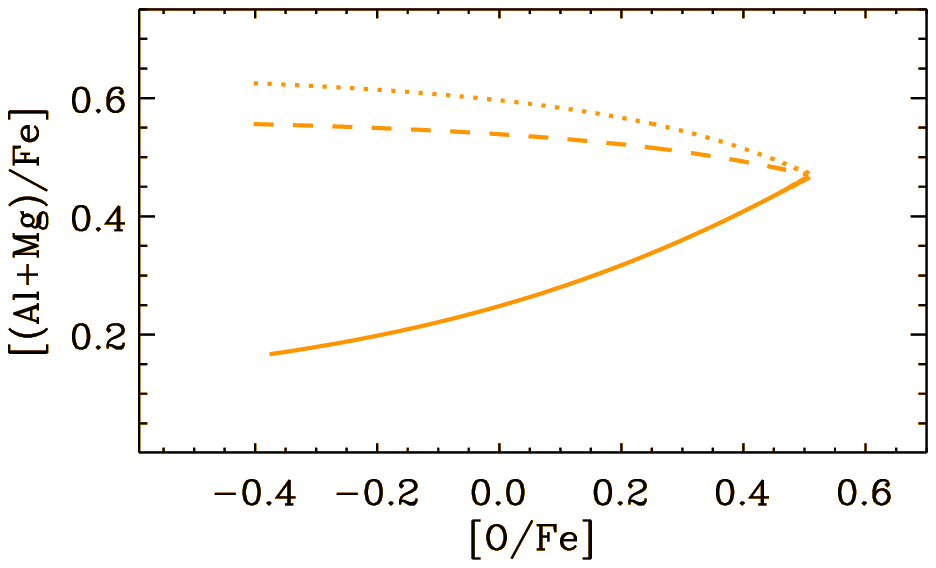 ,width=0.40\textwidth}
\end{center}   
\caption{Upper panel: the predicted evolution of the [Al/Fe] versus 
[Mg/Fe] abundance for NGC~6752. Also shown is the observational 
dataset of \citet{yong2005}. The solid line refers to the model 
with canonical Al yields from \citet{karakas2007} while the 
dashed line refers to our assumed mean Al production 
(Al=$3.5 \times 10^3$ M$_\odot$) from AGB stars that is required to reproduce 
the observations. The dotted line refers to the previous model 
but with a 50\% enhancement of the $^{25}$Mg and $^{26}$Mg yields 
from AGB stars (see Section~\ref{sec:mg_iso} for more details 
of this model).  Bottom panel: the predicted evolution of the 
sum [(Al+Mg)/Fe] versus [O/Fe] for the three models described 
above; line styles are the same.}
\label{fig:almg} 
\end{figure}

\begin{table} 
\centering 
\begin{minipage}{80mm} 
\caption{Isotopic ratios for the mean yields of SNe II, SNe Ia, and
AGB models. Notations are as in Tables \ref{tab:sne_mod} and
\ref{tab:agb_mod}.}
\label{tab:mod_iso} 
\begin{tabular} {|l|c|c|c|} 
\hline 
         & $^{25}$Mg/$^{24}$Mg  & $^{26}$Mg/$^{24}$Mg  & $^{12}$C/$^{13}$C    \\
\hline
SNe II (W\&W)  & 7.2 $\times 10^{-3}$ & 6.3 $\times 10^{-3}$ &       10$^{4}$      \\ 

SNe II (KOB)   & 9.9 $\times 10^{-3}$ & 8.9 $\times 10^{-3}$ &       10$^{3}$      \\

SNe II (C\&L)  & 2.8 $\times 10^{-3}$ & 5.0 $\times 10^{-3}$ & 5$\times 10^{4}$    \\
\hline
Model  (SNe II) &   $10^{-2}$          &  $10^{-2}$           &       10$^{4}$     \\     
\hline 
AGB     (KA07)& 6.5                  &     20.7             &         6.0         \\
\hline 
\end{tabular}   
\end{minipage}
\end{table}

\begin{figure*}    
\begin{center}    
\psfig{figure=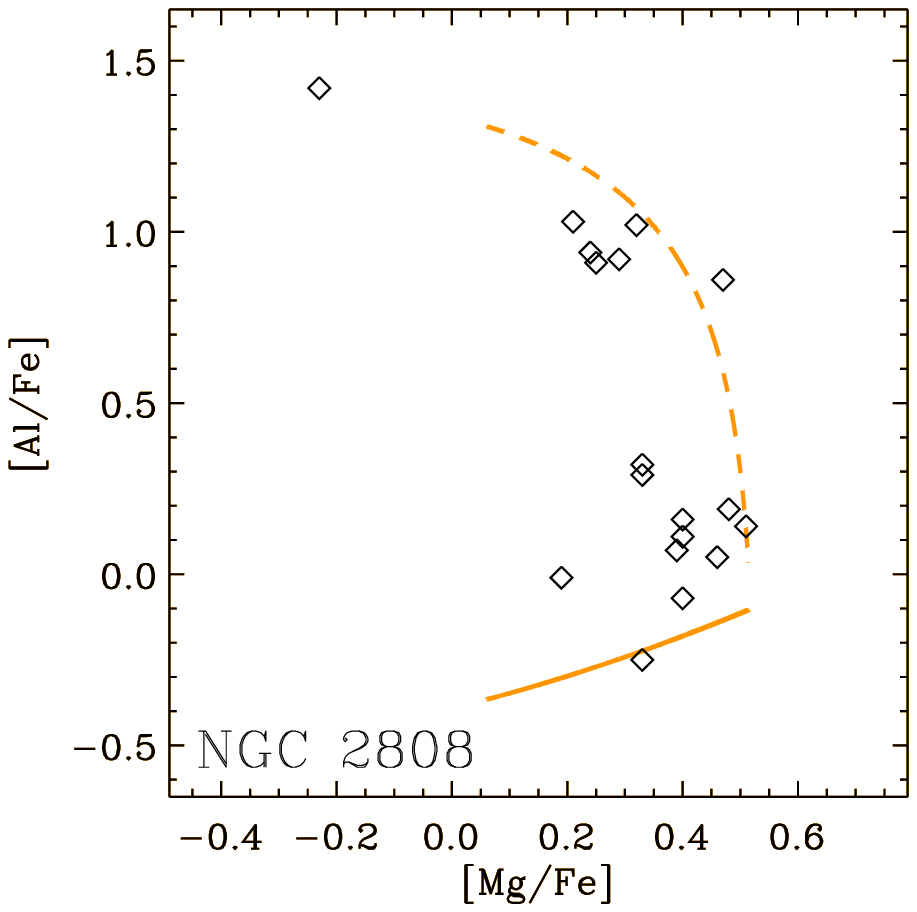 ,width=0.40\textwidth} 
\psfig{figure=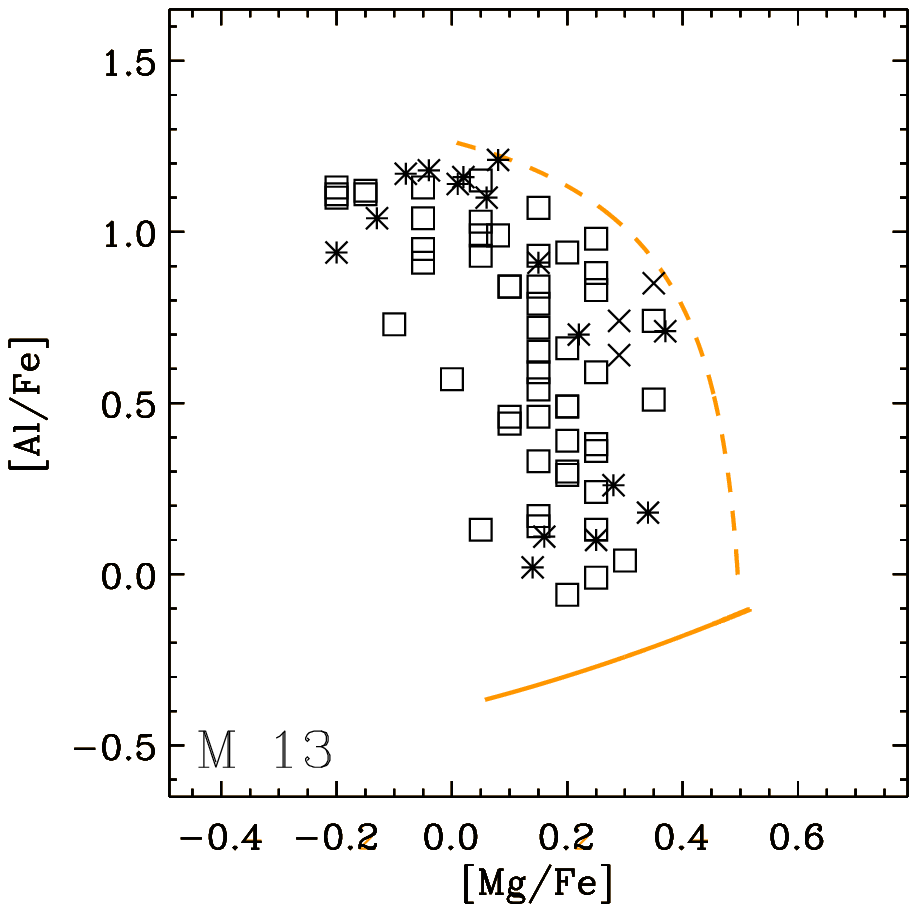 ,width=0.40\textwidth}
\psfig{figure=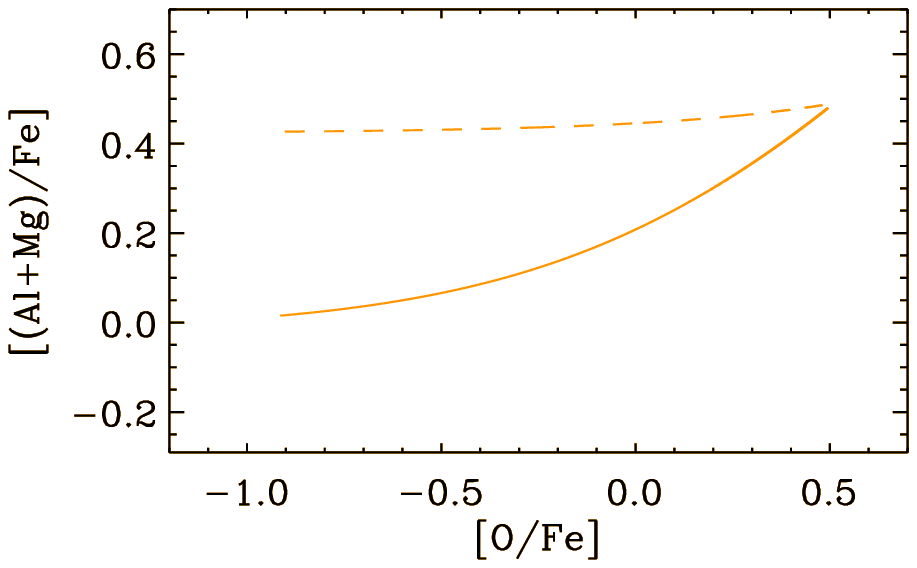 ,width=0.40\textwidth} 
\psfig{figure=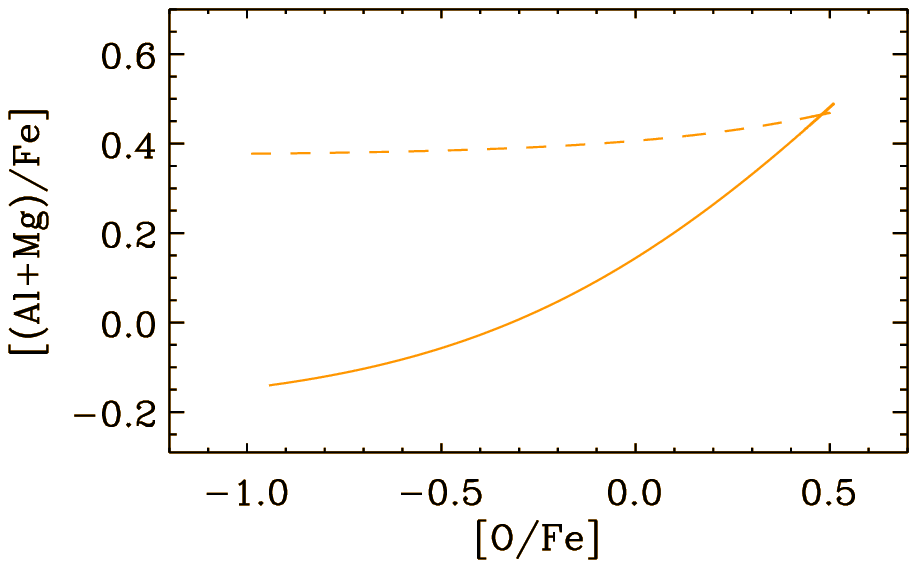 ,width=0.40\textwidth} 
\end{center}   
\caption{Upper panels: the evolution of the [Al/Fe] versus [Mg/Fe] abundance 
for NGC~2808 (left panel) and M~13 (right panel). As in
Fig.~\ref{fig:almg}, the solid lines refer to the model using yields
from \citet{karakas2007}, whereas the dashed lines refer
to the model with increased Al from AGB stars.  The evolution is
plotted against the observational datasets of \citet{carretta2006} for
the case of NGC~2808 (left panel) and \citet[][asterisks]{sneden2004},
\citet[][squares]{johnson2005} and \citet[][crosses]{cohen2005} for the
case of M~13. Regarding M~13, we point out that there is an
offset of 0.2 dex for the Mg abundance between our model and the
observational dataset. Bottom panels: the predicted [(Al+Mg)/Fe] 
versus [O/Fe] abundances; line styles are as above.}
\label{fig:almg_bis} 
\end{figure*}

\subsection{The Mg-Al anti-correlation}
\label{sec:almg}

In the left panel of Fig.~\ref{fig:almg} we compare the predictions
of our reference model (solid line) for NGC~6752 with the observed
Al-Mg anti-correlation found by \citet{yong2005}. This dataset shows
anti-correlations in stars that are both brighter and fainter than the
first giant branch bump, strongly suggesting that the peculiarities 
should be present in the gas from which the stars formed, and not caused
by internal stellar evolution. It is apparent that there is a 
 spread of $\sim 1.3$~dex in Al relative to Fe but only a modest spread of 
$\sim 0.3$~dex in Mg relative to Fe (the corresponding [O/Fe] spread 
discussed in Section~\ref{sec:nao} is $\sim 1.0$~dex).

In our framework, the first stars to form have enhanced Al
([Al/Fe]$\simeq1.4$) and depleted Mg ([Mg/Fe]$\simeq0.2$), while
subsequent generations evolve toward [Al/Fe]$\simeq$ 0.0 and
[Mg/Fe]$\simeq$0.5. It is apparent that our reference model
(Fig.~\ref{fig:almg}) can only account for the Mg variation and fails
to reproduce the $\sim 1.5$~dex spread in Al. The small amount of Al
produced in the models of \citet{karakas2007} is not enough to
countervail the large amount of Fe deposited by the single SN Ia:
the net results is that the [Al/Fe] ratio actually decreases in our
model. Note that even in the case of pure AGB pollution
\citep{fenner2004}, the maximum [Al/Fe] value is underestimated by a
factor of $\sim 0.7$~dex. If we assume that low-metallicity AGB stars
produce a mean value of Al$\sim 3.5 \times10^{-3}$ M$_{\odot}$
(instead of the literature values of $\sim 10^{-4}$ M$_{\odot}$) shown
in Table~\ref{tab:agb_mod}, the Al-Mg anti-correlation is remarkably
well reproduced.

In the upper panels of  Fig.~\ref{fig:almg_bis}, we  plot the [Mg/Fe]
versus [Al/Fe] evolution of our reference models for M~13 (right
panel) and NGC~2808 (left panel) against the corresponding
observational datasets \citep{sneden2004, johnson2005, cohen2005,
carretta2006}. Again the reference models are not able to reproduce
the Al enhancement, while models with a factor of $\sim50$ more Al
production from intermediate-mass AGB stars are able to reproduce the 
anti-correlation perfectly. The dotted line in the right panel represents 
the same dashed line but offset by 0.2 dex in [Mg/Fe]. At low [Al/Fe] 
values there is a difference of $\sim$0.2 dex between the [Mg/Fe] 
values of the two GCs.

At this point, it is worth asking if our proposed Al yield ($\sim 3.5
\times10^{-3}$ M$_{\odot}$) from intermediate-mass AGB stars is 
consistent with the uncertainties present in the theoretical models,
as well as with other available observations in other systems.
To test this latter point, we ran a test chemical evolution model 
of the Milky Way using GEtool \citep{fenner2002} with the same 
enhancement in Al AGB yields that is required for the GC model. 
The result is only a small difference (within $\sim 0.2-0.3$ dex 
and only for [Fe/H]$\ge$-1.6) compared with the canonical MW model.

The yields of $^{27}$Al are affected by variations in the nuclear 
reaction rates, as well as the treatment of mass loss and convection. 
For example both, 
the $^{26}$Mg(p,$\gamma$)$^{27}$Al and $^{26}$Al(p,$\gamma$)$^{27}$Si 
reactions play an important role in determining the final $^{27}$Al
abundance in a given AGB model.  \citet{izzard2007} concluded 
that the yields of $^{26}$Al and $^{27}$Al can vary by up to 
$\sim 2$ orders of magnitude, depending on the mass, metallicity
and choice of Mg-Al reaction rates. The yields of Al decrease with
increasing mass loss, because more mass is lost before significant
Al can be synthesized from Mg via the HBB. The yields of nitrogen and Al 
(but also C, Na and Mg) depend also on the third 
dredge-up \citep{ventura2005a,ventura2005b,karakas2006c}. While there
is uncertainty over the amount of dredge-up in intermediate-mass AGB
stars, observations suggest that it should be fairly efficient
\citep{wood83,garcia06}.  Moreover, hot-bottom burning is the major
production site for Al (as well as for Na and Mg).
In Fig.~\ref{fig:almg}  we show that increasing the AGB yields of Mg
by 50\% (particularly $^{25}$Mg and $^{26}$Mg, see next subsection) 
does not have a significant effect on the ability of the model to
reproduce the Mg-Al anti-correlation observed in NGC~6752. It
is not clear, however, if the model uncertainties would allow for
an increase in the Al abundance by a factor of 50 and further study
of this problem is required.

Using data for 18 stars, \citet{carretta2007} noted that the sum of 
Mg$+$Al is approximately constant in NGC~2808 for stars in all 
evolutionary phases (that is, constant over the whole magnitude 
range). The conclusion to be drawn from this observation is that
there has been a reshuffling of Mg into Al in these stars.
The bottom panels of Fig.~\ref{fig:almg} and Fig.~\ref{fig:almg_bis} show 
the sum of Mg+Al of our models plotted against the [O/Fe] abundance. 
If we assume the above mentioned Al enhancement in AGB yields, the sum 
Al+Mg appears to be surprisingly constant (in the reference model this 
sum is only constant within a factor of two). Increasing the yields of
Mg by 50\% does not change this result, and the sum of Mg$+$Al is
constant to within 0.1~dex (dotted line).

\begin{figure*}    
\begin{center}    
\psfig{figure=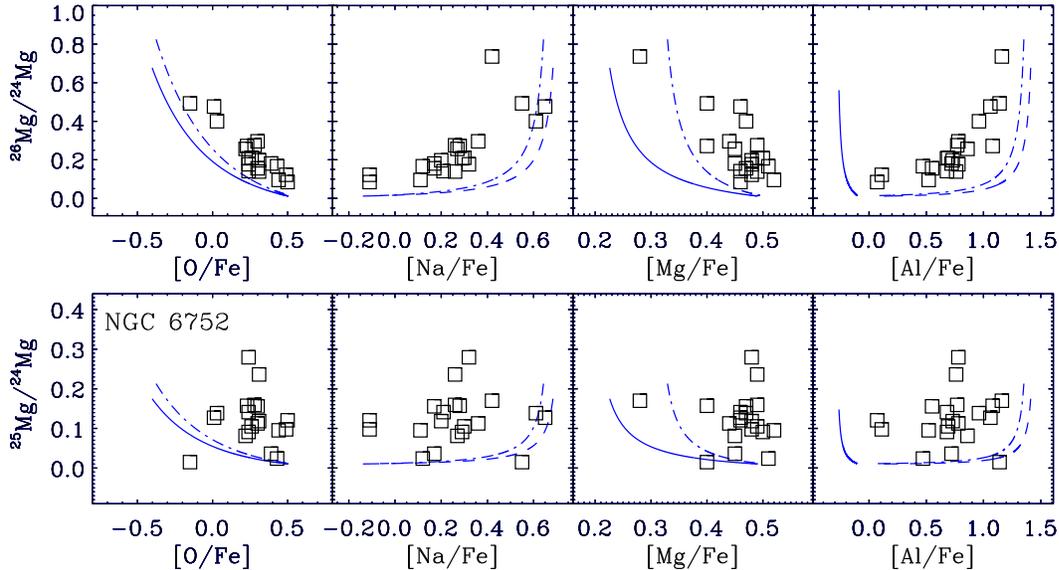,width=0.90\textwidth}  
\end{center}   
\caption{Predicted evolution of the Mg isotopic ratios versus [O/Fe],
[Na/Fe], [Mg/Fe], and [Al/Fe] for our model (dashed lines). The dashed
lines refers to our assumed ``Model'' (see Table~\ref{tab:agb_mod}),
while the dotted lines represent the evolution in the case of the 
$^{25}$Mg and $^{26}$Mg AGB yields increased by 50\%.
(see  Table~\ref{tab:mod_iso}).}
\label{fig:mg_iso} 
\end{figure*}

\subsection{The Mg isotope ratios}
\label{sec:mg_iso}

A further important test for our model is determining how well it
can reproduce the Mg isotopic ratios observed in GC stars
\citep[e.g. for NGC~6752][]{yong2003}.  The Mg
production from AGB stars is mainly in the form of the neutron-rich
magnesium isotopes, $^{25}$Mg and $^{26}$Mg, which are produced 
in the He-burning shell \citep[e.g.][]{karakas2003}. In
Table~\ref{tab:mod_iso} we summarize the mean $^{25}$Mg/$^{24}$Mg and
$^{26}$Mg/$^{24}$Mg ratios for the intermediate-mass AGB yields of
\citet{karakas2007}, as well as the same ratios for different SNe II
models computed by different authors. It is apparent that the ratios
$^{25}$Mg/$^{24}$Mg and $^{26}$Mg/$^{24}$Mg are much higher in the
case of AGB pollution compared to SNe II, where SNe II preferentially
produce $^{24}$Mg at low metallicities.  In the following we will
adopt a constant value of 0.01 for both $^{25}$Mg/$^{24}$Mg and
$^{26}$Mg/$^{24}$Mg to reflect SNe II (see Table~\ref{tab:mod_iso};
adopting values from different authors does not significantly change
the results).

In Fig.~\ref{fig:mg_iso} we show the observed Mg isotopic ratios 
for NGC~6752 from \citet{yong2003}, plotted against the 
[O/Fe], [Na/Fe], [Mg/Fe], and [Al/Fe]. In this figure we include
the predictions from our reference model (solid lines). The 
agreement is satisfactory, except in the case of Al (solid line) 
which, as already discussed, can be accommodated by assuming a 
larger Al production from AGB stars (dashed line). 
Both the $^{25}$Mg/$^{24}$Mg and $^{26}$Mg/$^{24}$Mg ratios decrease 
as the GC stars evolve from the AGB-SN Ia peculiar pollution 
($^{25}$Mg/$^{24}$Mg$\sim$0.5 and $^{26}$Mg/$^{24}$Mg$\sim0.15$), to 
predominantly self pollution by SNe II 
($^{25}$Mg/$^{24}$Mg$\sim$0.01 and $^{26}$Mg/$^{24}$Mg$\sim0.01$).

All GC chemical evolution models so far have failed to reproduce the
Mg isotopic ratios observed in NGC~6752 \citep[e.g.][]{fenner2004,
ventura2005c}. One of the strongest conclusions from \citet{fenner2004}
was that it is not possible to reproduce the Mg-Al anti-correlation
and the Mg isotopic ratios with a classical AGB self pollution
scenario. This is because AGB stars $produce$ Mg instead of destroying
it, and thus Mg and Al are correlated. In our model the [Mg/Fe] 
depletion is mainly caused by the effect of the SN Ia, hence 
the production of Mg from AGB stars is not significant. The Mg 
production instead helps to explain why strong [O/Fe] depletions 
(of up to $-0.5$) are not accompanied by a similarly low
[Mg/Fe] abundances. Even an increase of 50\% in the $^{25}$Mg and
$^{26}$Mg yields from AGB models does not effect our conclusions,
and instead meliorates the Mg isotopic ratios when plotted against
[Mg/Fe] (see Fig.~\ref{fig:mg_iso}).

\begin{figure}    
\begin{center}    
\psfig{figure=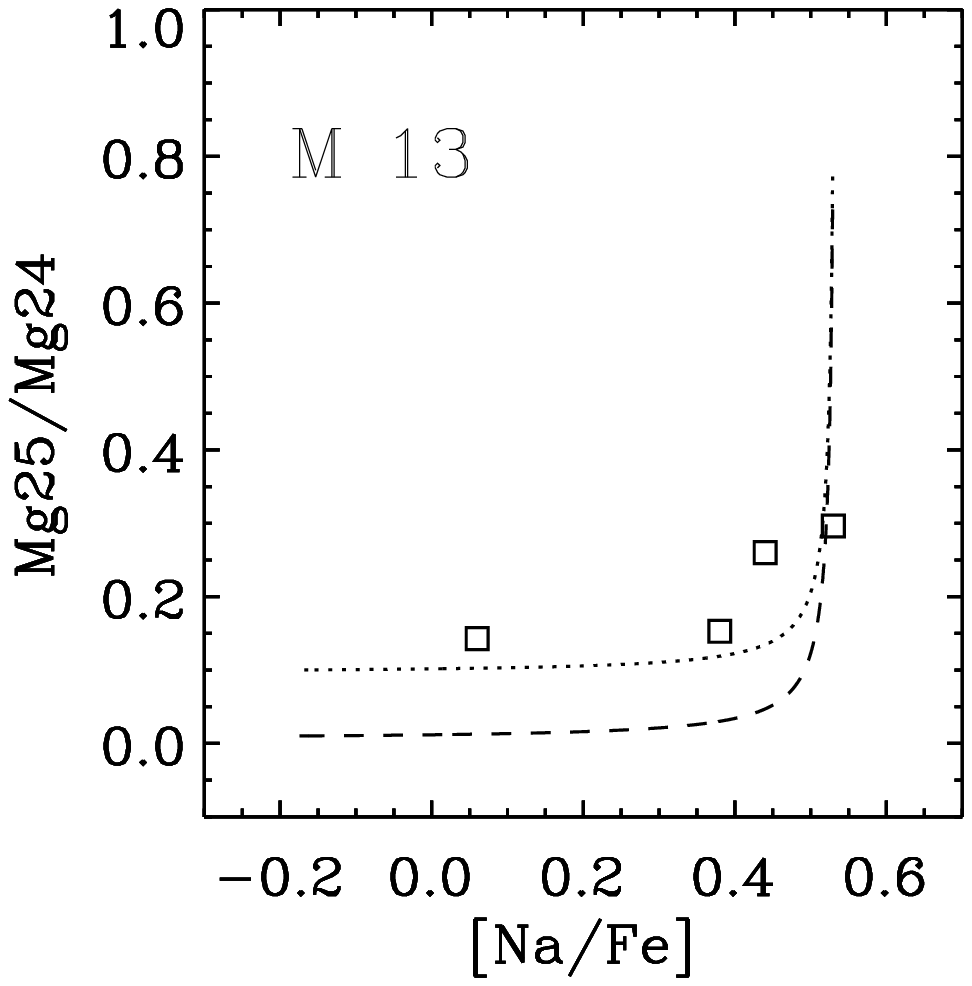,width=0.23\textwidth}  
\psfig{figure=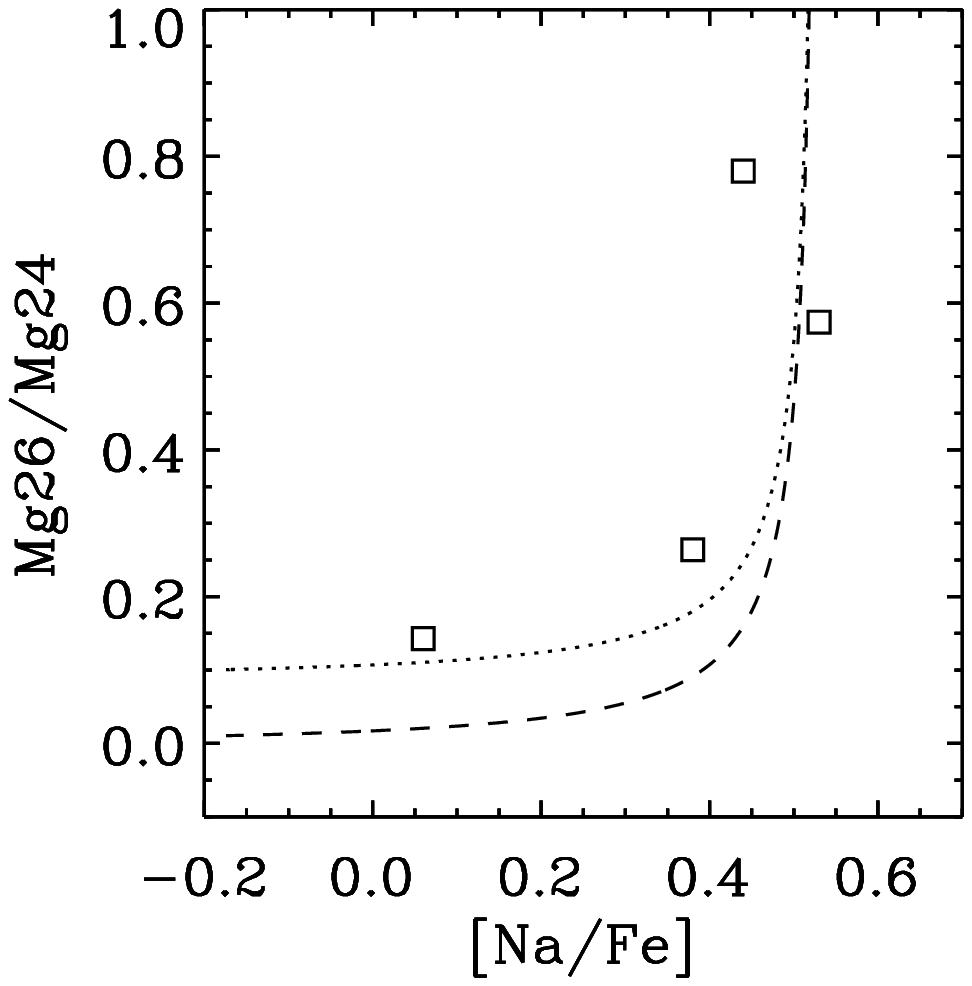,width=0.23\textwidth}
\end{center}   
\caption{Predicted evolution of Mg isotopic ratios versus [Na/Fe] for
our reference model (solid lines) compared with the observational
data from \citet{yong2006b}. The dashed lines correspond
to the model with $^{25}$Mg/$^{24}$Mg=0.1 and $^{26}$Mg/$^{24}$Mg=0.1 
from SNe II (instead of 0.01; see Table~\ref{tab:mod_iso}).}
\label{fig:mg_iso_bis} 
\end{figure}

Note that varying the Mg ratios in the SNe II yields from
$^{25}$Mg/$^{24}$Mg and $^{26}$Mg/$^{24}$Mg $\sim 0.01$ to 0.1 would
help to fit the observational data.  In Fig.~\ref{fig:mg_iso_bis} we
show the Mg isotopes ratios versus [Na/Fe] from \citet{yong2006b} in
the case of M~13 together with the predictions from our model and, again, 
the observations are matched fairly well.  Assuming
$^{25}$Mg/$^{24}$Mg=0.1 and $^{26}$Mg/$^{24}$Mg=0.1 for the SNe II
yields does an even better job at reproducing the observations. 

\subsection{The C-N anti-correlation}
\label{sec:cn}

\begin{figure}    
\begin{center} 
\psfig{figure=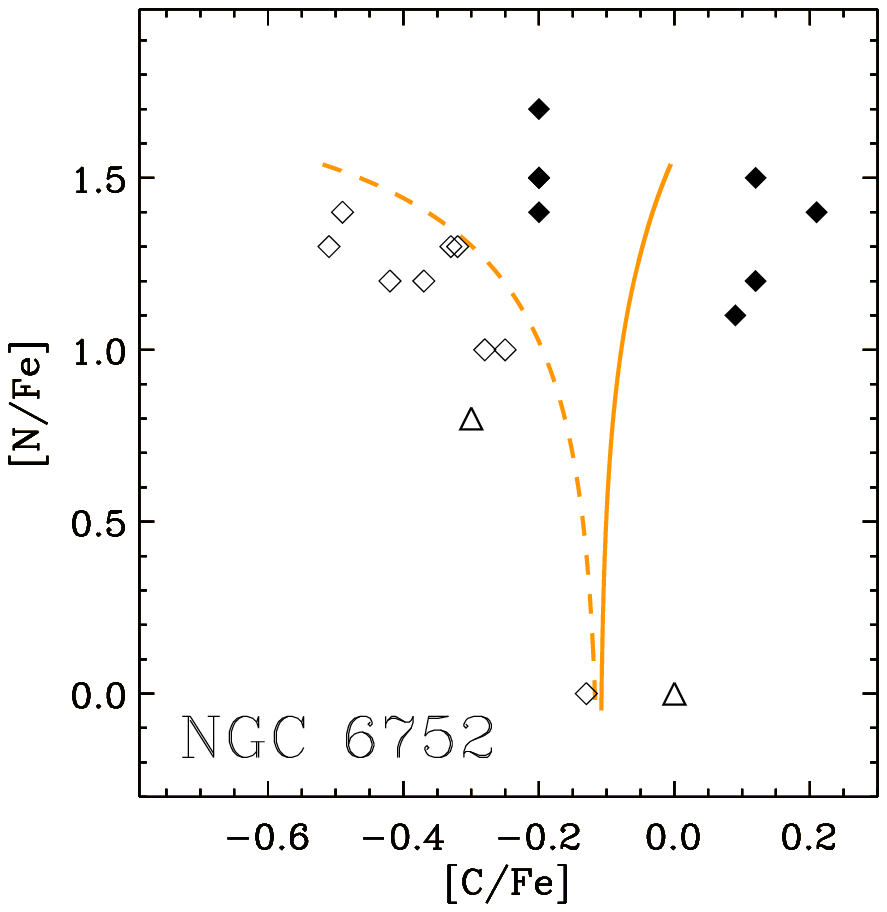,width=0.40\textwidth}    
\psfig{figure=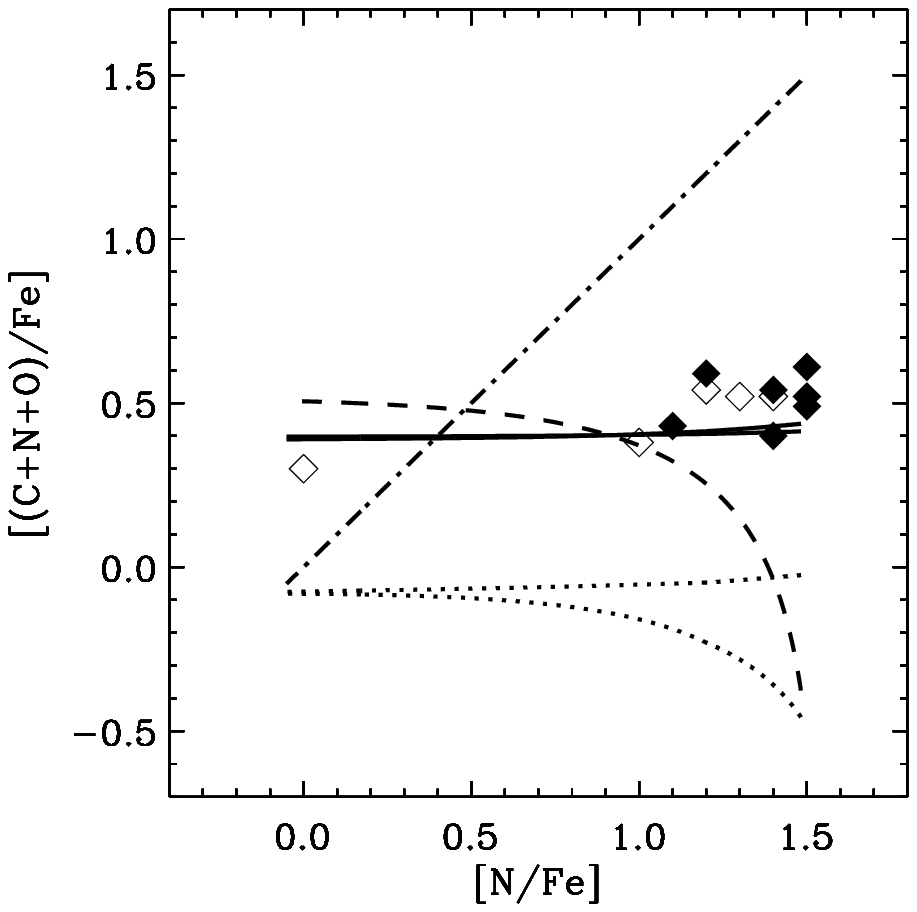,width=0.40\textwidth}    
\end{center}   
\caption{Upper Panel: evolution of [C/Fe] versus [N/Fe] for the
reference model (solid line) for NGC~6752, plotted against
observational data of two AGB stars by \citet{smith1993} (triangle),
and 9 sub-giants stars (open diamonds) and 7 dwarf stars (filled
diamonds) by \citet{carretta2005} (squares). The dashed line
represents the model with AGB C yields decreased by a factor of 4.
Lower panel: evolution of the sum of [(C+N+O)/Fe] (solid lines) for
the reference model (thick lines) and well as for the model with
reduced C (thin lines) plotted against the observational dataset of
\citet{carretta2005}; the symbols are the same as in the upper panel. 
The evolution of each single element is also shown: 
C (dotted lines), N (dot dashed line), and O (dashed line).}
\label{fig:cn}  
\end{figure}

In Fig.~\ref{fig:cn} we illustrate the evolution of [C/Fe] versus
[N/Fe] of our reference model for NGC~6752, together with
observational data by \citet{carretta2005} for 9 dwarfs (filled
diamonds), 9 sub-giants stars (open diamonds), along with 2 giants
stars (open triangles) observed by \citet{smith1993}. Comparison with
stars near the main sequence is preferred, because variations due to
the first dredge-up and deep mixing can affect the abundances of
evolved stars. However, it is very difficult to obtain high-quality
spectra of main-sequence stars owing to their low brightness. In the
following we will not discuss the effect of stellar evolution on
the changes to the surface abundances of C and N \citep[instead we refer
the interested reader to][and references therein]{smith1992,
charbonnel1994,charbonnel1995, denissenkov2000, weiss2000,
gratton2004}.

While the more evolved stars (open symbols) show a clear C-N
anti-correlation, similar to that observed in other GCs (see also
Fig.~\ref{fig:cn_bis} for M~13), the inclusion of the dwarf stars
makes it difficult to distinguish a clear evolutionary pattern.
\citet{carretta2005} pointed out that the spectra for these turn-off
stars did not have enough quality to allow for an  accurate
determination of the carbon abundances; the same also applies to the 
$^{12}$C/$^{13}$C ratios (see \S~\ref{sec:c_iso}). Typical errors for 
the [C/Fe] abundances are of the order of $\sim 0.2$~dex with the 
exception of the dwarf stars that show larger uncertainties.

A quick examination of Table~\ref{tab:agb_mod} shows that nitrogen is
the element produced in the largest proportions by intermediate-mass AGB
stars.  Thus, it is not surprising that the first stars forming in the
inner region initially polluted by AGB stars show very high values of
[N/Fe] ($\sim1.5$) despite the action of the SN Ia. For comparison, in
\citet{fenner2004}, stars polluted with AGB ejecta show [N/Fe] values
as high as $\sim$2.0). Carbon is also produced in intermediate-mass
AGB stars, although not to the same level as N, hence the [C/Fe]
ratios remains low. A glance at Fig.~\ref{fig:cn} shows that while a
large variation of [N/Fe] is achieved in our model in accordance with
observations, the ratio of [C/Fe] remains almost constant. While
[C/Fe]$\sim0.0$ at high values of [N/Fe] is consistent with the
observed [C/Fe] abundances inferred from dwarf stars, assuming a
factor of 4 less carbon is produced by AGB stars does a much better 
job of reproducing the C-N anti-correlation observed in sub-giant stars.
(see Fig.~\ref{fig:cn} and Table~\ref{tab:agb_mod}).

The C-N anti-correlation is also seen for stars on the main sequence
turn off of M~13 \citep{briley2004a} (see Fig.~\ref{fig:cn_bis}),
whereas we are not aware of a similar dataset for NGC~2808 (although
the C-N evolution of this GC is similar to M~13 for more evolved
stars).  We refer to \citet{briley2004a} for a discussion of the
uncertainties affecting the abundances; these can be large (up to
0.5~dex, see their Fig.~5) owing to the faintness of the observed
stars.  Again the two models show quite similar results, with [C/Fe]
remaining approximately constant in the reference model, and the model
that assumes C is under produced by a factor of 4 doing the best job
of reproducing the C-N anti-correlation.

To test the possible contribution of low-metallicity fast rotators or
mass loss in metal-poor massive stars \citep{meynet2002a,meynet2002b}
we qualitatively increase the nitrogen mean yield for SNe II, from our
reference value of [N/Fe]=$-0.90$ to values of [N/Fe]=$-0.30$ and 0.0.
This increase only slightly changes the results, by increasing the
minimum value of [N/Fe]$\simeq -0.4$ in Fig.~\ref{fig:cn_bis} by a 
factor of 0.20-0.40 dex.

\subsection{The sum of C$+$N$+$O}
\label{sec:cno}

One of the most important observational constraints that any GC 
chemical evolution model needs to fulfill is the constancy of the sum C$+$N$+$O
\citep[e.g][]{ivans1999, carretta2005}. Self pollution scenarios
using AGB models calculated using the mixing-length theory (MLT) of convection
\citep{fenner2004, karakas2006} predict that stars forming from different
generations of AGB ejecta show large increases in the CNO sum. This
is in contrast to AGB models computed using the ``full spectrum of turbulence'' 
convective model \citep{ventura2005b} that manage to keep the sum of CNO 
constant within a factor of two. This is because the \citet{ventura2005b}
show little third dredge-up (which increases C and hence N) and more 
efficient HBB (which converts C and O to N).  Hence, MLT models create 
C and N rich AGB ejecta at almost constant O 
\citep[see Fig. 4 in][]{fenner2004}. FST models result in moderately 
O-depleted AGB ejecta that is only moderately enriched in N 
\citet{ventura2005b}

\begin{figure}    
\begin{center} 
\psfig{figure=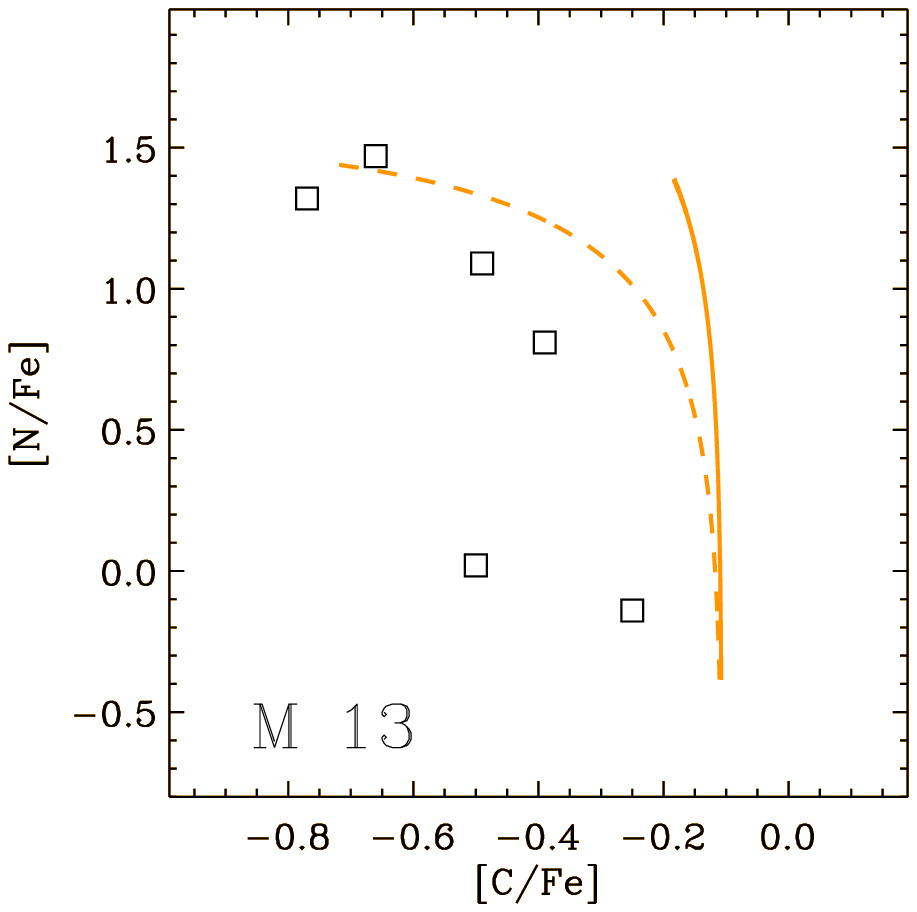,width=0.43\textwidth} 
\psfig{figure=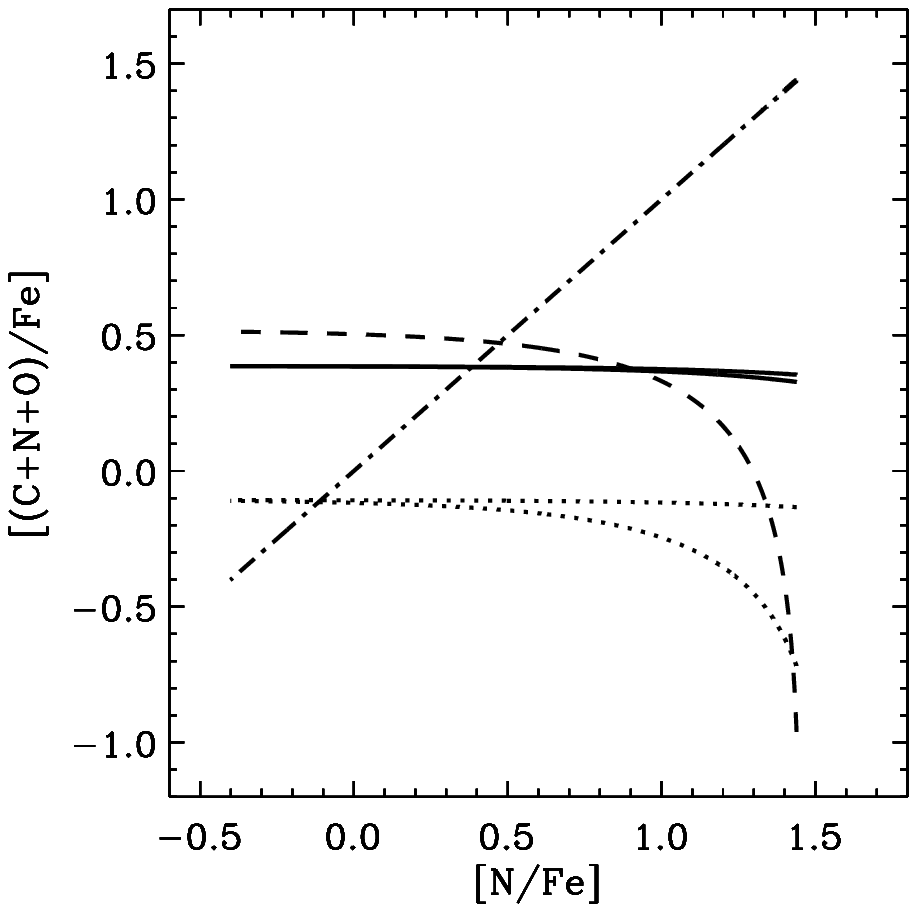,width=0.43\textwidth} 
\end{center}   
\caption{M~13 Upper Panel: evolution of [C/Fe] versus [N/Fe] for
the reference model (solid line) plotted against the observational 
data of \citet{briley2004b}. The dashed line corresponds to the 
model where we decrease the C yields from AGB stars by a factor of
4. Lower panel: the evolution of the sum of [(C+N+O)/Fe] (solid lines) 
for the reference model (thick lines) as well as for the model with
reduced C (thin lines). The evolution of each single element is also 
shown: C (dotted lines), N (dot dashed line) and O (dashed line). 
The corresponding two diagrams for the case of NGC~2808 are almost
identical to the results for M~13 and not shown.}
\label{fig:cn_bis} 
\end{figure}

In the bottom panel of Fig.~\ref{fig:cn} we show the C$+$N$+$O sum
(solid line) of our model for NGC~6752, together with [C/Fe] (dotted
lines), [N/Fe] (dot-dashed line) and [O/Fe] (dashed line) plotted
against [N/Fe]. In the same figure we also include the observational
data from \citet{carretta2005} for 5 sub-giants and 7 dwarf stars
from the same GC. It is quite remarkable that our model predicts a
practically constant CNO sum and is a satisfactory fit to the
observations.  It is apparent that while the [C/Fe] abundance remains
approximately constant, a large variation of [N/Fe] comes with a similarly
large variation in [O/Fe]. The N deposited by AGB stars is
counteracted by the O production in SNe II when the inner region expands
and mixes (lowering the [N/Fe] value]). In principle, in our model,
there is no reason for this to happen because the mechanisms controlling
the Fe abundance and the CNO sum are different (i.e., SNe II, SN Ia,
and AGB).  For example, more AGB pollution (or no AGB
pollution at all) would increase (decrease) the initial CNO sum and
the condition C+N+O$=$constant would not be fulfilled. However, in the
three particular GC considered here, once the free parameters are
fixed to reproduce the observed anti-correlations, a constant C+N+O
naturally arises.

The thin solid lines and dotted lines in the lower panels of
Fig.~\ref{fig:cn} and Fig.~\ref{fig:cn_bis} refer to the case in which
the C production from AGB stars is lowered by a factor of 4 to better
fit the C-N anti-correlation. In this case the sum of C$+$N$+$O also
remains quite constant, in agreement with the observations. The fact
that a variation of 0.4~dex of the [C/Fe] abundance only slightly
changes the CNO sum is understandable if we consider that O and N are
a factor of $\sim$ 8 more abundant than C at low values of [N/Fe] and
high values of [N/Fe], respectively.

\subsection{The carbon isotope ratio}
\label{sec:c_iso}

\begin{figure}    
\begin{center} 
\psfig{figure=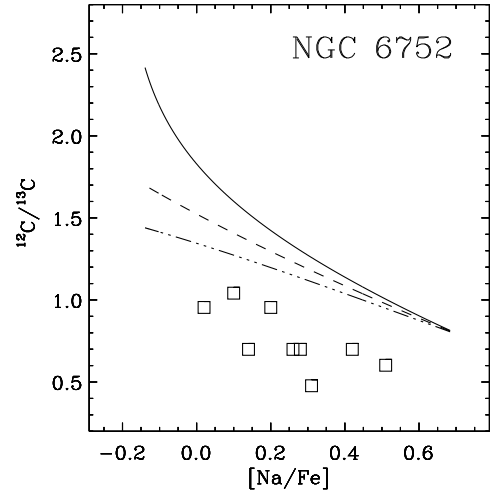,width=0.40\textwidth}    
\end{center}   
\caption{Carbon isotopic ratio $^{12}$C/$^{13}$C versus [Na/Fe] for
the NGC~6752 reference model, plotted against the observational
dataset of \citet{yong2003}. The dashed and dot-dashed lines represent
a model in which the possible contribution of low metallicity, 
fast rotators are qualitatively taken into account according to
\citet{chiappini2008}.}
\label{fig:cna_iso} 
\end{figure}

In Fig.~\ref{fig:cna_iso} we show the evolution of the
$^{12}$C/$^{13}$C ratio versus [Na/Fe] from our model plotted against
observational data from \citet{carretta2005} for sub-giant stars in
NGC~6752.  Owing to the poor quality of the spectra, these authors were
not able to derive the carbon isotopic ratio for dwarf stars. 
Our model over-predicts the $^{12}$C/$^{13}$C ratio at any value
of [Na/Fe] and by up to a factor of 1.5 dex. \citet{chiappini2008} 
and \citet{hughes08} note
that a model without fast rotators gives very high values of the
$^{12}$C/$^{13}$C ratio of up to $10^{4}$ because SNe II mainly produce
$^{12}$C (solid line in Fig.~\ref{fig:cna_iso}). Such high values 
are not reached in Fig.~\ref{fig:cna_iso} because of the partial 
contribution from intermediate-mass AGB stars, that also produce low
$^{12}$C/$^{13}$C ratio yields (see Table~\ref{tab:mod_iso}).
The \citet{chiappini2008} model with 
rapidly rotating massive stars brings the ratio down to 
$^{12}$C/$^{13}$C$\sim$80 (dashed line) at [Fe/H]=$-3.5$ and down to
$^{12}$C/$^{13}$C$\sim$30 (dot-dashed line) at [Fe/H]=$-5.0$. 
If low-metallicity, rapidly rotating massive stars contribute some 
$^{13}$C, the predicted carbon isotope ratio will be lower but not 
as low as required to match the observational data (see dashed 
and dot-dashed line in Fig.~\ref{fig:cna_iso}).

Evidently it is not possible to reproduce the observed
$^{12}$C/$^{13}$C ratios without invoking internal stellar evolution
(e.g., the first dredge-up and/or extra-mixing processes). In the
 more luminous sub-giant stars the observed isotopic ratio
$^{12}$C/$^{13}$C is very low, typically in the range of $\sim 3$ to
11 (noting that the equilibrium value of the CNO cycle is $\approx 3$).
The first dredge-up can reduce the ratio from 90 (solar) to $\sim 20$ in
standard stellar evolution models \citep[e.g.,][]{boothroyd1999}.
\citet{shetrone2003} found that in NGC~6528 and M~4 this ratio
declines steeply with increasing luminosity along the RGB. The
discontinuity in the $^{12}$C/$^{13}$C ratio occurs at the bump
luminosity and has also been detected in metal-poor field halo 
stars \citep{gratton2004}.

\citet{carretta2005} inferred a lower limit of 10 for the 
$^{12}$C/$^{13}$C ratio for the three dwarf stars in NGC~104. 
Even if these values are more in accordance with our model, further
determinations of the $^{12}$C/$^{13}$C ratio in turn-off stars is
needed before making a more detailed comparison.


\subsection{The Helium Content}
\label{sec:he}

The unusual HB morphology observed in NGC~2808, which exhibits an
extended blue tail and a gap separating the red and the blue clumps
\citep{bedin2000}, can be reproduced if the blue stars have a higher
helium content ($Y\sim0.32$) compared to those in the red clump, with
primordial helium ($Y\sim0.24$) \citep{dantona2004}. 
NGC~2808 is also known to have a peculiar main sequence 
\citep{dantona2005}, in which the bluer stars are inferred to
have an higher helium content from fitting theoretical isochrones to
the observed data. To reproduce the colour-magnitude diagram (CMD),
\citet{dantona2005} suggested that $\sim$ 20\% of the stars
have a helium content as high as $Y \sim 0.40$, $\sim$ 30\%  have 
a spread of $Y$ between 0.24 and 0.29, while the remaining 50\% of 
the stellar population have primordial $Y$. Further observations 
discovered  a triple main sequence in NGC~2808 \citep{piotto2005}, 
that has given support to the extreme value of $Y \sim 0.40$ required 
to fit the CMD.

Even if such large $Y$ values are required to reproduce the main features
of the CMD, the origin of the helium is not known. Helium values
as high as $Y=0.40$ are very difficult to reproduce with 
canonical chemical evolution models without violating other 
observational constraints \citep[e.g.][]{karakas2006}. \citet{dantona2004} 
required a factor of $\sim 10$ more intermediate-mass AGB stars 
than what it is inferred from a Salpeter-like IMF to return the 
amount of $Y$ needed to form the number of blue HB stars in NGC~2808. 
\citet{karakas2006} highlights the difficulty in producing the 
large postulated helium enrichment using the AGB self-pollution 
scenario. In particular, their largest predicted value of 
$Y\sim0.29$ is accompanied by a large increase in sum C$+$N$+$O.

\begin{figure}    
\begin{center}    
\psfig{figure=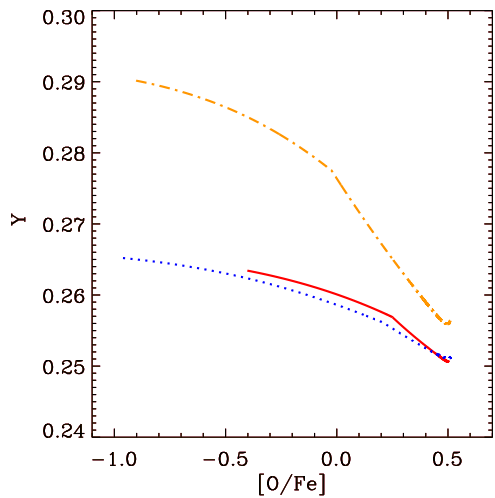,width=0.40\textwidth} 
\end{center}   
\caption{Evolution of the helium content, $Y$, versus [O/Fe]. 
The red solid line represents the model for NGC~6752; the blue 
dotted line for M~13; the dot dashed orange line for NGC~2808.}
\label{fig:he} 
\end{figure}

In Section~\ref{sec:cno} we show that one of the main successes of
this model is the ability to reproduce the C$+$N$+$O sum for all three
GCs NGC~6752, M~13, and NGC~2808. In Fig.~\ref{fig:he} we plot the
helium content $Y$ of the reference models for each cluster against
their [O/Fe] abundances. Note that we assume a primordial value of
$Y\sim0.248$ \citep{coc2004}. It is quite evident that in the cases of
NGC~6752 and M~13 the increase is quite modest and within $\Delta Y
\sim 0.015$, whereas for the specific case of NGC~2808 we obtained
$Y\sim 0.29$. This larger rise in the $Y$ values is due to the larger
density of polluting intermediate-mass AGB stars in NGC~2808, which is
$\sim$2-3 times greater than for the other GCs.

Even if NGC~2808 is the only GC out of the three that shows a
appreciable helium variation, a maximum value of $Y \sim 0.29$ is
smaller than the postulated very high value of $Y \sim 0.40$ required
to explain the most enriched stars in the triple main sequence
\citep{piotto2007}. It is very close, however, to the value of 
$Y \sim0.32$ \citep{bedin2000} required to explain the unusual 
HB morphology for NGC~2808.

\section{Comparison with previous models and general comments}

 To explain the peculiar chemical properties observed in GCs,
we propose a model based on a new idea of 
pre-peculiar inhomogeneous pollution produced
by the explosion of a {\it single} SNe~Ia.
In principle, due to the low star formation rate during the formation of the
proto-halo, the possibility that 2 or 3 (or more) SNe~Ia explode
essentially simultaneously ($<$20-30~Myr apart in time) in a region smaller than
30-50 pc is much less probable than having a single
event. Having said that, our model can accomodate such a 
scenario without a dramatic impact on the predictions based upon
the base scenario of a single SNe~Ia event.
For example, if 2 (or 3) SNe Ia explode in the same region,
ejecting therefore 
a factor of 2 (or 3) more iron, an expansion (i.e., an increase of R$_{\rm in}$)
of a factor of $\sqrt[3]{2}$ (or $\sqrt[3]{3}$)
would lead to the same
[Fe/H] and chemical properties seen in the base model.
In the case
of no SNe Ia, but intermediate-mass stellar AGB pollution, one is left with the
classical ``external'' GC self enrichment scenario \citep{bekki2007}
in which the high values of N, Na (and other elements produced in
substantial amounts by AGB
stars) are easily explained, but it is very difficult to explain the
depletion in oxygen. Since all the well-observed GCs seem to have an
oxygen-depleted population of stars at some level, in our framework
we must assume that the inhomogenous contribution of SN Ia is the {\it seed}
for the formation of GCs.

Of course, a larger number of SNe Ia may contribute to the enrichment
of the proto-halo ISM (outside the proto-GC region).
We believe, though,  that the approximation of an ISM being enriched
only by SNII products is sound one, as the SNe~Ia 
contribution to the ``mean'' MW's stars'
chemical properties starts being important only after $\sim$1.5 Gyr and at
[Fe/H] $> -1$ (Tinsley 1979; Edvasrdsson et al. 1993; McWilliam 1997).
The [Fe/H]$_{\rm ISM}$ values in our models are, in all cases,
lower than this value.

\citet{bekki2007} point out that one of the greatest advantages
of an external AGB-pollution scenario is that the fraction of peculiar
stars observed in GCs is not limited to the amount of a stars forming
in the first generation. In particular, in this model the fraction of
O-depleted to O-rich stars depends only in the way that the SNe II
pollution proceeds, while the number of AGB stars initially polluting
the inner region is quite low ($N_{\rm AGB}\simeq 200$; see
Table~\ref{tab:gc_mod}). The total amount of stars needed to create a
typical value of 200 intermediate-mass AGB stars is $\sim
10^{4}$ M$_{\odot}$, which is a small fraction of the total MW Halo or
GC mass. Even assuming that this number of stars are trapped in the
potential well of the forming GC and are homogeneously
distributed within 50 (or 100) pc, their density would be
$1.9\times10^{-2}$ ($2.4\times 10^{-3}$) M$_{\odot}$pc$^{-3}$ which is
low compared to the densities observed in GCs
\citep[e.g.,][]{pryor1993}.

In principle, our proposed scenario has no particular problems
explaining the observed high fraction of peculiar to normal stars
observed in some GCs \citep[e.g.,][for NGC~2808]{carretta2006}. The
self-pollution scenario, on the other hand, requires ad hoc
assumptions such as an anomalous IMF peaked at 4--8 M$_{\odot}$
\citep[e.g.,][]{smith1982, dantona2004, prantzos2006}; or that the GC
was initially much more massive and has lost $\sim$90-99 per cent of
the stars that were formed in the first generation through tidal
stripping. Our model does not require these assumptions: the only
assumption we require is that star formation should proceed more or
less smoothly while SNe II explode and self enrich the proto-GC; and
the chemical properties of the forming stars evolve from peculiar to
normal. The fraction of normal to peculiar stars depends on how
the SF proceeded and we will try to address this problem with detailed
hydrodynamical simulations.
 
In
previous models 
\citep[e.g.][]{fenner2004, dantona2005, bekki2007} the CN bi-modality
is not reproduced due to the lack of C-depleted and N-rich stars,
since C is produced in AGB stars. The same is also true for Mg, with
the result that positive C-N and Mg-Al correlations are found in
contradiction to the observational constraints. \citet{bekki2007}
tried to solve this problem by using yields from AGB models that had
no third dredge-up (as these authors noted, this assumption is purely
hypothetical). The smaller production of C is what allows this study
to reproduce the C-N anti-correlation. The O-Na and Al-Mg trends were
not recovered because the lack of dredge-up resulted in a smaller
production of Na and Mg (mainly $^{26}$Mg).  Note that in all AGB
models, O destruction via HBB is not efficient enough to account for
the very low O abundances observed in some GC stars (typically [O/Fe]
$\approx -0.5$ or less).  The non-standard models of
\citet{ventura2005c} \citep[see also][]{ventura2006} that use a
different convection model, can reproduce most of the abundance
anomalies but only when including convective overshooting to force
some third dredge-up along with adopting high mass-loss rates. 
In our model none of the above assumptions are required.  
We use the AGB yields of \citet{karakas2007}, averaged over a 
canonical Salpeter IMF.

The success of our model is the combined effect of inhomogeneous AGB
pollution {\it together with} the effect of a single SN Ia and its
Fe-rich ejecta. The net effect is to lower the ratio [$X_{i}$/Fe],
where the element $X_{i}$ is negligibly produced (or destroyed)
relative to SN II in intermediate-mass AGB stars (e.g., oxygen). In
the case where $X_{i}$ is produced by AGB stars, the ratio is either
maintained at an approximately constant value or the enhancement is
mitigated owing to the effect of the SN pollution (e.g., C, Na, Mg and
possibly Al). The only real problem that our model suffers is the
production of Al should be enhanced by a factor of $\sim 10$--50
compared to the AGB yields from \citet{karakas2007}. Our model,
however, is the first that has shown any promise at explaining the Mg
isotopic ratios in GC stars.
 
One possibility in a canonical AGB self-pollution model is that the
second generation of stars did not form from AGB-enriched material but
these stars had their surfaces polluted by AGB winds
\citep[e.g.,][]{dantona1983}. In our framework the O-depleted stars
obtain their abundances from SNe Ia pollution mixed with AGB
material. We could image a similar surface pollution model as follows:
after the first ``normal'' generation of stars have formed, self
pollution by intermediate-mass AGB stars and SNe Ia (which has similar
timescales; see Section~\ref{sec:initial_conditions}) start
inhomogeneously polluting the surface of stars. It would be difficult,
however, to explain all the chemical properties of GC stars with
surface pollution. For example, it would probably not be possible to
maintain [Fe/H] constant in the most O-depleted stars. While surface
pollution of SN Ia ejecta would decrease the [O/Fe] ratio it would
also dramatically increase the surface Fe content of the star.

\section{Conclusions}
\label{sec:conclusion}
 
We have modelled the chemical evolution of GCs under the hypothesis
that the abundance anomalies were set during the formation of the
Milky Way Halo (or inside their host dwarf galaxies).  Inside a volume
of gas enriched to some level by a first generation of low-metallicity
SNe II, GC formation takes place owing to the inhomogeneous (respect
to the surrounding ISM) effect of a single SN Ia plus material from
intermediate-mass AGB stars.  After a burst of star formation, new SNe II
start to explode, self-polluting and expanding out the inhomogeneous
region, which mixes with the surrounding (metal poorer) ISM. In our
framework, all the peculiar stars observed in GCs (e.g., O-depleted,
Na-rich) were born in the above inhomogeneously enriched volume, while
``normal'' stars (i.e., O-rich, Na-depleted) are formed successively
once the SNe II associated with the SF have washed out the
inhomogeneous region and self-polluted the ISM. In conclusion, 
our main findings can be summarized as follows:

\begin{enumerate}

\item We reproduce the O-Na anti-correlation. In our model the
first stars to form have low [O/Fe] and high [Na/Fe] abundances,
with the system then evolving toward ``normal'' abundances once SNe II
start polluting the ISM.

\item We reproduce the Mg-Al anti-correlation, but only 
by assuming that intermediate-mass AGB produce more Al than 
predicted by theoretical models \citep[e.g.,][]{karakas2007}. This 
assumption needs to be carefully studied in the context of AGB model
uncertainties.

\item We satisfactorily produce the Mg isotopic ratios
observed in NGC~6752.

\item The [Fe/H] abundance remains approximately constant 
during the evolution. This is because of mixing caused by the subsequent 
SNe II explosions between the Fe-enriched inhomogeneous inner region 
and the [Fe/H]-poorer ISM.  The Fe produced by the SNe II themselves does
not significantly affect the resulting Fe abundance in the forming GC stars.

\item We reproduce the C-N anti-correlation within observational
uncertainties, and the sum C$+$N$+$O remains $\approx$ constant,
in agreement with the observations. The best fit is obtained, however,
by assuming that the AGB stars produce a factor of four less C. This
assumption may not be in accord with current theoretical models.

\item The model cannot reproduce the $^{12}$C/$^{13}$C ratios
observed in evolved giant stars in the GCs, and internal stellar 
evolution processes are required.

\item GCs are a factor of $\sim$ ten more metal rich than coeval 
MW halo stars (which are the assumed stellar host population).

\end{enumerate}

In a forthcoming paper we will focus on other elements (e.g, Li, F, Si,
Ca, Ti, V, Co, Ni, and Cu) to further test our framework. We will also
focus upon the extreme metallicity GC cases that may prove to be a 
challenge for our model: the very metal poor M15 ([Fe/H]=$-2.26$) 
and the metal-rich 47 Tuc ([Fe/H]=$-0.76$). Preliminary results are 
encouraging.

One last question we could ask is why would all GCs show such a
peculiar enrichment as outlined in this paper?  Even if we do not have
a conclusive answer, the region enriched by the SN Ia and AGB stars
has a [Fe/H] abundance and metallicity, $Z$, that is a factor of $\sim$
ten higher than the surrounding ISM.  Thus could be due to the corresponding
enhanced cooling, this region may be more likely than the halo field
to produce a star cluster with regions of enhanced star formation. In
our framework, the SN Ia (and AGB) pollution is the seed which may
have led to the formation of the globular cluster itself.

\section*{Acknowledgments}

We thank the anonymous referee whose comments improved the
presentation of the paper.  We also thank Eugenio Carretta and
Maurizio Salaris for useful discussions regarding GC chemical
abundances and isochrone, respectively.  It is also a
pleasure to thank Annibale D'Ercole for reading the paper and
providing useful suggestions which improved the presentation of the
paper.  BKG acknowledges the support of the UK's Science \& Technology
Facilities Council (ST/F002432/1) and the Commonwealth Cosmology
Initiative.  AIK thanks Onno Pols for useful discussions and
acknowledges support from the Australian Research Council's Discovery
Projects funding scheme (DP0664105).  PSB is supported by a Marie
Curie Intra-European Fellowship within the 6th European Community
Framework Programme.

\bibliographystyle{mn2e}    
\bibliography{globular_refs}    
   
\label{lastpage}    
\end{document}